\documentclass[a4paper,fleqn]{cas-dc}
\usepackage[justification=centering]{caption}
\usepackage{bm}
\usepackage[authoryear,longnamesfirst]{natbib}

\usepackage[longnamesfirst, authoryear]{natbib}  

\usepackage[justification=centering]{caption}
\usepackage[mathscr]{eucal}

\usepackage{graphicx}

\def\tsc#1{\csdef{#1}{\textsc{\lowercase{#1}}\xspace}}
\tsc{WGM}
\tsc{QE}

\begin{document}
\let\WriteBookmarks\relax
\def\floatpagepagefraction{1}
\def\textpagefraction{.001}

\shorttitle{\title{Multi-Level Sequence Denoising with Cross-Signal Contrastive Learning for Sequential Recommendation}}
\shortauthors{Zhu et~al.}

\title[mode = title]{Multi-Level Sequence Denoising with Cross-Signal Contrastive Learning for Sequential Recommendation}




\author[1]{\textcolor{black}{Xiaofei Zhu}}
\ead{zxf@cqut.edu.cn}
\credit{Conceptualization of this study, Methodology, Writing - original draft \& review \& editing, Supervision}
\cormark[1]

\author[1]{\textcolor{black}{Liang Li}}
\ead{liliang@stu.cqut.edu.cn}
\credit{Conceptualization of this study, Methodology, Software, Writing - original draft}

\author[2]{\textcolor{black}{Weidong Liu}}
\ead{liuweidong@chinamobile.com}
\credit{Supervision}

\author[3]{\textcolor{black}{Xin Luo}} 
\ead{luoxin@swu.edu.cn}
\credit{Writing - review \& editing, Supervision}

\address[1]{\label{address:cqut}College of Computer Science and Engineering,  Chongqing University of Technology, Chongqing 400054, China}
\address[2]{China Mobile Research Institute, China Mobile , Beijing 100053, China}
\address[3]{College of Computer and Information Science, Southwest University, Chongqing 400715, China}

\cortext[cor1]{Corresponding author}




\begin{abstract}
Sequential recommender systems (SRSs) aim to suggest next item  for a user based on her historical interaction sequences. Recently, many research efforts have been devoted to attenuate the influence of  noisy items in sequences by either assigning them with lower attention weights or discarding them directly. The major limitation of these methods is that the former would still prone to overfit noisy items while the latter may overlook informative items. To the end, in this paper, we propose a novel model named \underline{M}ulti-level \underline{S}equence \underline{D}enoising with \underline{C}ross-signal \underline{C}ontrastive \underline{L}earning (MSDCCL) for sequential recommendation. To be specific, we first introduce a target-aware user interest extractor to simultaneously capture users' long and short term interest  with the guidance of target items. Then, we develop a multi-level sequence denoising module to alleviate the impact of  noisy items by employing both soft and hard signal denoising strategies. Additionally, we extend existing curriculum learning by simulating the learning pattern of human beings. It is worth noting that our proposed model can be seamlessly integrated with a majority of existing recommendation models and significantly boost their effectiveness. 
Experimental studies on five public datasets are conducted and the results demonstrate that the proposed MSDCCL is superior to the state-of-the-art baselines. 
The source code is publicly available at https://github.com/lalunex/MSDCCL/tree/main.
\end{abstract}


\maketitle

\section{Introduction}
{S}{equential} Recommender Systems (SRS) have emerged as a challenging research field and play a critical role in present e-commerce and social media platforms \citep{xie2022contrastive,chen2023data,wang2023poisoning}, with the aim of recommending the next item based on users' historical sequences. 
The core problem of SRS is how to effectively capture the sequential patterns of historical user behaviors. To this end, many kinds of methods based on deep neural networks have been proposed.  For example, some works propose to leverage recurrent neural networks (RNNs) \citep{hidasi2015session, orvieto2023resurrecting}, convolutional neural networks (CNNs) \citep{tang2018personalized,sudarsan2023selfattention}, Transformer \citep{ma2023improving}, and Graph Neural Networks (GNNs) \citep{hao2023multi} to model the sequential patterns. Despite these methods have achieved promising performance, they have a major limitation that  user sequences may contain some noisy items (e.g., misclick \citep{tolomei2019you} and malicious false interactions \citep{zhang2020practical}, which undoubtedly pose significant challenges for extracting user preference and making recommendations.  

In recent years, many research efforts have been devoted to handle the noisy issue in users' historical sequences. One research line is to mitigate the impact of noise within sequences in a ``soft'' manner by utilizing the attention mechanism \citep{li2020time, li2021hyperbolic} or filtering algorithms \citep{zhou2022filter}. These methods  attempt to reduce the influence of noisy items by assigning them lower attention weights when learning representations of users' historical sequences.  DSAN \citep{yuan2021dual} introduces a virtual target item and takes it as the query vector for attention weight assignment. AC-TSR \citep{zhou2023attention} introduces two calibrators, i.e., a spatial calibrator and an adversarial calibrator, to assign the attention weights via utilizing spatial relationships and each item's contribution for prediction. 
FMLP-Rec \citep{zhou2022filter} applies fast fourier transform to convert items into the frequency domain and utilize a low-pass filter to alleviate  noisy items in the frequency domain. 
Although these research efforts have achieved encouraging results,  
they still suffer from noisy items in sequences \citep{zhang2022hierarchical}.
Another research line proposes to explicitly remove noisy item in a ``hard'' manner \citep{sun2021does,zhang2022hierarchical}. 
BERD \citep{sun2021does} proposes to handle unreliable items in sequences by modeling their corresponding losses and uncertainties via a Gaussian distribution.
HSD \citep{zhang2022hierarchical} combines two levels of signals, i.e., the user-level signal and the sequence-level signal, to identify inherent noisy items in sequences. 
One major limitation of these methods is that they would ignore relevant items in sequences during the denoising process, which will lead to inferior performance.

To address the issues mentioned above, we propose a novel method called Multi-level Sequence Denoising with Cross-signal Contrastive Learning (MSDCCL) for sequential recommendation. The main idea of our model is to handle noisy items in user interaction sequences by simultaneously leveraging both ``soft'' and  ``hard'' denoising strategies, where each of them mutually guide the learning of the other. 
Specifically, we develop a soft- and hard-level sequence denoising module which consists of two sub-modules, i.e., a soft-level denoising sub-module and a hard-level denoising sub-module. The former alleviates noisy items by assigning lower attention weights, and the latter eliminates irrelevant items identified by using a Gumbel-Softmax function \citep{chaudhary2023gumbel,wang2023multi}. It is worth noting that we also introduce a cross-signal contrastive learning layer to allow guidance informative exchange between the two sub-modules. To the best of our knowledge, this is the first model that can comprehensively explore the benefits of both ``soft'' and  ``hard'' denoising strategies for sequential recommendation.  
In addition, we incorporate a target-aware user interest extractor to  model long- and short-term user interests, where a transformer sub-module is utilized to capture the long-term user interest and a target-aware convolutional sequence embedding sub-module is developed to learn effective short-term user interest.  
At last, we  extend existing
curriculum learning \citep{wang2023curriculum,xu2024curriculum} by utilizing a S-shape function to simulate the learning process of human beings. To be specific, we separate ``difficult'' training samples into low- and high-speed learning zones, and train our model with fewer samples in the low-speed learning zone and more samples in the high-speed learning zone.





Extensive experiments have been conducted to examine the performance of our proposed model MSDCCL on five  public benchmark datasets, i.e., ML-100k, Beauty, Sports, Yelp and ML-1M. The results demonstrate that MSDCCL significantly outperforms the state-of-the-art methods. For example, the relative performance improvements of MSDCCL over the two best performing baselines (i.e., HSD+BERT4Rec and AC-BERT4Rec) are 100.00\% and 233.96\% in terms of HR@5 on the dataset ML-100k, respectively.  We further conduct an ablation study to verify the contribution of each component in our model, and the results show that removing each of them will lead to a considerable performance degradation. At last, we  investigate the influence of the S-shape curriculum learning, and the results suggest that MSDCCL equipped with S-shape increment performs better than its conterpart equipped with linear increment. 
We summarize  the main contributions of this paper as follows:
\begin{itemize}
    \item  We propose a novel denoising model for the task of sequential recommendation, which simultaneously attenuates the denoising issue in the sequence via both soft and hard denoising strategies. 
    \item  We design a target-aware user interest extractor to effectively capture both user long-term and short-term interest.
    \item   We extend existing curriculum learning by simulating the learning pattern of human beings.
    \item   Extensive experiments have been conducted on five widely used datasets, including ML-100k, ML-1M, Beauty, Sports and Yelp. The results show that the proposed  MSDCCL is superior to all state-of-the-art baseline methods. 
\end{itemize}

The rest of the paper is organized as follows. Section 2 gives the related work. Section 3 provides the preliminaries. We introduce the details of our proposed approach MSDCCL in Section 4, and present the experimental results and analysis in Section 5. Section 6 draws the conclusions of this paper.

\begin{figure*}[htbp]  
      \centering  
      \includegraphics[width=17cm,keepaspectratio]{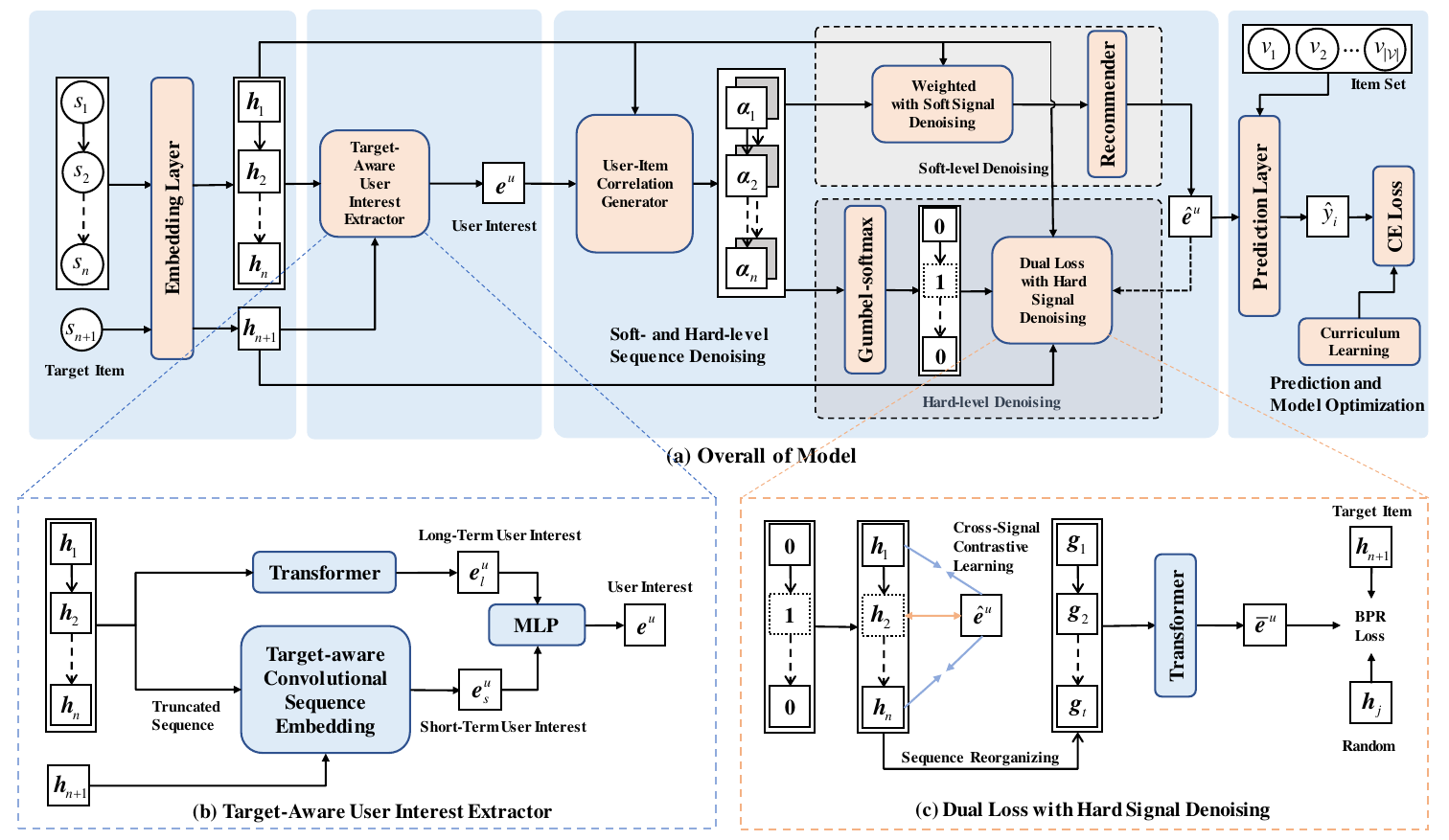} 
      \caption{The architecture of our proposed model MSDCCL.} 
      \label{fig:overall_framework} 
\end{figure*}

\section{Related Work}   
\subsection{Sequential Recommendation}
Sequential recommendation aims at extracting a user's interest from her historical interactions and then predicting the next item that she is most interested in. Some pioneer works \citep{rendle2010factorizing} utilize Markov Chains (MCs) to capture dynamic changes in user interest. 
With the tremendous success of deep learning, in recent years, researchers  resort to developing various deep neural networks based methods for sequential recommendation and have achieved encouraging progress. 
Hidasi et al. (\citeyear{hidasi2015session}) 
first propose to employ recurrent neural networks (RNNs) for sequential recommendation and introduce a novel ranking loss function for model training.
Liu et al. (\citeyear{liu2016context}) model user sequential behaviors by leveraging the contextual information. They extend the conventional RNN models by introducing variable input matrix and transition matrix.
Tang et al. (\citeyear{tang2018personalized}) utilize convolutional filters to extract sequential patterns and develop a convolutional sequence embedding recommendation model which assigns larger impact to more recent items.  
Yuan et al. (\citeyear{yuan2019simple}) extract both short and long-range item dependency to learn high-level representation, and increase the receptive fields by employing a stack of holed convolutional layers. 
Wu et al. (\citeyear{wu2019session}) model session sequences as graph-structured data and apply graph neural networks (GNNs) to capture complex transition patterns between items. 
Wang et al. (\citeyear{wang2020make}) further consider both the transition information and the corresponding temporal dynamics  of items to learn accurate representations. They  utilize external knowledge graph to capture the relation between items, and introduce a dynamic time kernel function to  merge these representations.

\subsection{Contrastive Learning}
Contrastive Learning (CL)  \citep{huang2022self} has been successfully used in the field of recommendation, which attempts to extract useful self-supervised signals from user behavior sequences. Many efforts have been conducted to apply contrastive learning to improve the recommendation performance via augmenting the user-item bipartite graph. For example, 
Wu et al. (\citeyear{wu2021self}) utilize dropout-based structure perturbation to augment the user-item bipartite graph, and  maximize the representation consistency between same nodes in different augmented graphs. 
Different to the dropout-based augmentation, Yu et al. (\citeyear{yu2022are}) propose a noise-based augmentation by constructing contrastive views via incorporating different random noises to the original representations. They aim to regularize the embedding space towards a uniform distribution. 

Some other works attempt to apply contrastive learning  to yield better performance for sequential recommendation. For example, 
Xia et al. (\citeyear{xia2021self}) introduce the hypergraph representation learning to alleviate the sparse issue of the session data, and provide strong self-supervision signals for recommendation.
Qiu et al. (\citeyear{qiu2021memory}) develop a multi-instance noise contrastive estimation, which extends the noise contrastive estimation from a single positive sample to a multi-instance variant.
Xie et al. (\citeyear{xie2022contrastive}) employ stochastic data augmentation to transform each user interaction sequence into two correlated views of the sequence, and explore contrastive learning on user interaction sequences to boost sequential recommendation performance. 
Qiu et al. (\citeyear{qiu2022contrastive}) incorporate the target item to provide supervision signals for positive sampling, and design a contrastive regularization to improve the representation distribution.
Chen et al. (\citeyear{chen2022intent}) leverage unlabeled user behavior sequences to capture users' intent distribution, and propose an intent contrastive learning module to maintain the agreement between a sequence   and its corresponding intent. 
To reduce the false positive and false negative samples, Wang et al. (\citeyear{wang2022explanation}) employ explanation methods to derive positive and negative sequences based the importance of items in a user sequence. 
Duan et al. (\citeyear{duan2023clsprec}) leverage the same user's short-term and long-term interaction sequences as positive samples, and other randomly selected interaction sequences as negantive samples.
Zhu et al. (\citeyear{zhu2024high}) first learn a user's low-level and high-level preference interest, and then take the corresponding preference interest between them as positive samples, and others as negative samples.

The key difference between our method and existing models is that we propose a novel cross-signal  (i.e., the generated soft and hard supervision signals) to generate the positive and negative samples. Specifically, we first utilize the generated hard supervision signals to identify noisy and relevant items. Then, we take these pairs between the user interest representation (i.e., generated via the soft supervision signals) and the noisy items as negative samples, and pairs between it and the relevant items as positive samples.

\subsection{Denoising Methods}
Due to the existence of noise within users' historical interactions \citep{wang2023efficient,zhang2023denosing,quan2023robust}, e.g., accidental interactions, a surge of works have been proposed to deal with noisy items in sequences.  These methods can be roughly grouped into two categories, i.e., soft-denoising category and hard-denoising category. The methods in the soft-denoising category mainly depend the self-attention mechanism, which  attempts to assign lower attention weights to noisy items in order to alleviate their influence.
Li et al. (\citeyear{li2020time}) consider the influence of both the position  of each item as well as the time interval between items on prediction and propose a time-aware self-attention mechanism for recommendation. 
Luo et al. (\citeyear{luo2020collaborative}) first identify neighborhood sessions of the current session and then utilize the self-attention network to assign weights to collaborative items.   
Yuan et al. (\citeyear{yuan2021dual})  design a self-attention network to obtain target embedding and apply a vanilla attention network to estimate the importance of items in sequences. An adaptively sparse transformation function is then incorporated to eliminate the influence of noisy items in sequences.
Zhou et al. (\citeyear{zhou2023attention}) improve the estimation of attention weights for items in a sequence by developing a spatial calibrator and an adversarial calibrator. The former is designed to adjust attention weights based on spatial information among items, and the latter is used to modify attention weights by exploring the contribution of each item. 
However, these methods would still assign attention weights to less relevant or irrelevant items in sequences, leading to sub-optimal recommendation performance.

The methods in the hard-denoising category aim to explicitly drop noisy items in sequences.  
Tong et al. (\citeyear{tong2021pattern}) employ reinforcement learning to determine the relevance of each item in sequences. They investigate sequential patterns for the policy learning process and formulate the denoising problem as a Markov Decision Process.  
Chen et al. (\citeyear{chen2022denoising}) propose a denoising strategy by getting rid of noisy attentions. They introduce a trainable binary mask in each self-attention layer which  assigns zero attention scores to noisy items. This method remains the architectures of transformers, while changing the attention distributions.
Zhang et al. (\citeyear{zhang2022hierarchical}) propose a hierarchical sequence denoising method by introducing two types of signals, i.e., user-level intent signals and sequence-level context signals, to identify inconsistency items in  sequences. 
Despite the promising performance achieved by these methods, 
identifying irrelevant items in sequences is still challenging and relevant information would be discarded during the denoising process.
Different with previous works, we propose to combine both soft and hard denoising strategies in a unified model, where the two denoising strategies will guide the denoising process of each other.

\section{Preliminaries} 
In our denoising model, we have a collection of users and items, denoted as $\mathcal{U}$ and $\mathcal{V}$, correspondingly, where $u \in \mathcal{U}$ stands for a user and $v \in \mathcal{V}$ stands for an item. The quantity of users and items in the collection is indicated as $\vert\mathcal{U}\vert$ and $\vert\mathcal{V}\vert$, respectively. Our model takes the collection of users’ historical interactions $\mathcal{S}=\{S^{u_{1}}, S^{u_{2}}, \cdots, S^{u_{\vert\mathcal{U}\vert}}\}$ as input. For a particular user, her interaction sequence is represented as $S=[s_{1}, s_{2}, \cdots, s_{n}]$, where $\textit{n}$ indicates its length, and $s_{i}$ denotes the $\textit{i}$-th item that has been interacted. The aim of sequential recommendation is to predict the next item that the user is most likely to engage with at the $(n+1)$-th step, depicted as $p(s_{n+1}|s_{1:n})$.

\section{Methodology} 
The architecture of our proposed model MSDCCL is illustrated in Figure \ref{fig:overall_framework}. It mainly consists of four components: embedding layer, target-aware user interest extractor, soft- and hard-level sequence denoising, and prediction layer.

\subsection{Embedding Layer}
In the embedding layer, $\boldsymbol{M}_{I} \in \mathbb{R}^{|\mathcal{V}| \times d}$ denotes a dense, lower-dimensional item embedding matrix, where $\textit{d}$ represents the size of embedding vectors. The corresponding item embedding matrix $\boldsymbol{H} \in \mathbb{R}^{n \times d}$ for a user’s history interaction sequence $S$ can be obtained by performing a look-up operation on $\boldsymbol{M}_{I}$. To capture the temporal influence of each item in $S$, we also incorporate a learnable position embedding matrix $\boldsymbol{P} \in \mathbb{R}^{n \times d}$. Therefore, for an item $s_{i}$ in $S$, we have $\boldsymbol{h}_{i} \in \boldsymbol{H}$ and $\boldsymbol{p}_{i} \in \boldsymbol{P}$ to represent its item embedding and corresponding position embedding, respectively.  

\subsection{Target-Aware User Interest Extractor}
User historical interactions reflect the evolution of her interests, leading to different interests between long- and short-term interactions \citep{lv2020time,zheng2021cold}. To represent the long-term user interest, we utilize a transformer encoder to model the whole historical interactions. Previous research has emphasized the importance of modeling the short-term user interest based on her recent interactions. In this regard, it is common to utilize the last $\textit{m}$ items in the historical interactions to capture the short-term user interest \citep{liu2018stamp,zheng2022disentangling}. In contrast, some researches employ the next item to reflect the short-term user interest \citep{zhang2022hierarchical,lin2022sparse}. Due to potential shifts of the short-term user interest, we integrate those two approaches during the training phase to improve its stability. As the target item remains unobservable during the validation and testing phases, we exclusively use the last $\textit{m}$ items in the sequence to model the short-term user interest. After acquiring the representations of user’s long- and short-term interest, we utilize a feedforward neural network to fuse them into a comprehensive representation of user interest.

\subsubsection{Long-term User Interests} Inspired by the works of \citep{zhou2020s3,xu2023group}, our model incorporates a transformer encoder to encode users' historical interactions. It consists of multiple identical layers that comprise of multi-head self-attention and feed-forward neural networks, which can be formulated as follows:
\begin{equation}
    \boldsymbol{H}^{l}=\text{Transformer}\left(\boldsymbol{H}^{l-1}+\boldsymbol{P}\right), \quad \forall l \in[1, \cdots, L],
\end{equation}
where $\boldsymbol{H}$ and $\boldsymbol{P}$ represent the item embedding matrix and the corresponding position embedding matrix, respectively. To obtain the final output of the transformer encoder, we retrieve the hidden representation matrix $\boldsymbol{H}^{L}=\left[\boldsymbol{h}_{1}^{L}, \boldsymbol{h}_{2}^{L}, \cdots, \boldsymbol{h}_{n}^{L}\right] \in \mathbb{R}^{n \times d}$ from the last layer $L$. For simplicity, we utilize the hidden representation of last item $\boldsymbol{h}_{n}^{L}$ as the long-term user interest $\boldsymbol{e}_{l}^{u} \in \mathbb{R}^{d}$.

\begin{figure*}[]  
      \centering  
      \includegraphics[width=12cm,keepaspectratio]{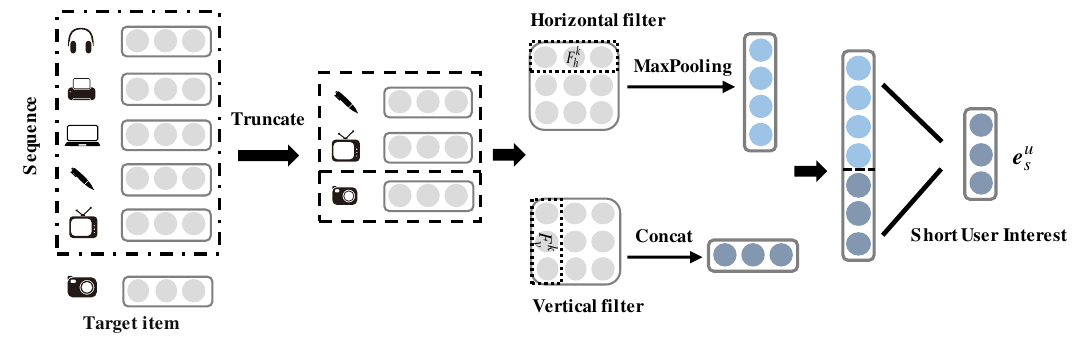} 
      \caption{Target-aware convolutional sequence embedding. } 
      \label{fig:convolution} 
\end{figure*}

\subsubsection{Short-term User Interests}
 Based on previous discussions, we propose to integrate both the last $m$ items and the next item to learn a high quality short-term user interest representation. This integration entails jointly modeling the actual target item and the last $\textit{m}$ items, which is formulated as follows:
 \begin{align} 
        \boldsymbol{H}^{\prime} &= \operatorname{Truncate}(\boldsymbol{H}, m), \\
        \boldsymbol{c}_{h}^{k} &= \operatorname{Conv}_{h}^{k}\left(\left[\boldsymbol{H}^{\prime} \| \boldsymbol{h}_{n+1}\right], \boldsymbol{F}_{h}^{k}\right), \forall k \in [1, \cdots, z(m+1)], \\
        \boldsymbol{c}_{v}^{k} &= \operatorname{Conv}_{v}^{k}\left(\left[\boldsymbol{H}^{\prime} \| \boldsymbol{h}_{n+1}\right], \boldsymbol{F}_{v}^{k}\right), \forall k \in\left[1, \cdots, z^{\prime}\right], 
\end{align}
where $\operatorname{Truncate(\cdot)}$ is a function which truncates the last $\textit{m}$ items from the sequence $S$, $\boldsymbol{h}_{n+1}$ is the target item representation, and $\|$ denotes the concatenation operation. It's important to note that $\boldsymbol{c}_{h}^{k}$ and $\boldsymbol{c}_{v}^{k}$ are obtained by applying convolution with a stride of 1, where $\boldsymbol{F}_{h}^{k}$ and $\boldsymbol{F}_{v}^{k}$ stand the horizontal and vertical convolution kernel, respectively. Meanwhile, $z$ and $z^{\prime}$ represent the number of convolution kernels of the same shape. The specific convolution process is shown in Figure \ref{fig:convolution}. Subsequently, we fuse $\boldsymbol{c}_{h} \in \mathbb{R}^{z\left(m+1\right)}$ and $\boldsymbol{c}_{v} \in \mathbb{R}^{z^{\prime}d}$ to obtain the short-term user interest as follows:
\begin{align}
    \boldsymbol{c}_{h} &= \operatorname{MaxPooling} \left(\operatorname{ReLU}\left(\boldsymbol{c}_{h}^{1}, \boldsymbol{c}_{h}^{2}, \cdots, \boldsymbol{c}_{h}^{z\left(m+1\right)}\right)\right),\\
    \boldsymbol{c}_{v} &= \left[\boldsymbol{c}_{v}^{1} \| \boldsymbol{c}_{v}^{2} \| \cdots  \| 
    \boldsymbol{c}_{v}^{z^{\prime}}\right], \\
    \boldsymbol{e}_{s}^{u} &= \operatorname{MLP}\left(\left[\boldsymbol{c}_{h} \| \boldsymbol{c}_{v}\right]\right).
\end{align}

\subsubsection{Interests Fusion}
After obtaining the long- and short-term interests of user, we employ a feedforward network to combine the two interest representations, yielding the final user interest $\boldsymbol{e}^{u} \in \mathbb{R}^{d}$, which is formulated as follows:
\begin{align}
    \boldsymbol{e}^{u} = \left(\operatorname{ReLU}\left(\boldsymbol{e}_{s}^{u}\boldsymbol{W}_{1} + \boldsymbol{e}_{l}^{u}\boldsymbol{W}_{1}\right)\right)\boldsymbol{W}_{2}+ \boldsymbol{b}_{1}
\end{align}
where $\boldsymbol{W}_{1}, \boldsymbol{W}_{2} \in \mathbb{R}^{d \times d}$ and $\boldsymbol{b}_{1} \in \mathbb{R}^{d}$ denote the trainable parameters.

\subsection{Soft- and Hard-level Sequence Denoising}
\subsubsection{User-Item Correlation Generator}
Users' historical interaction sequences inevitably contain noise \citep{tolomei2019you} and in some cases, even malicious false interactions \citep{ zhang2020practical}. The noise within interaction sequences can potentially mislead the model, resulting in sub-optimal user representations. To identify noisy items in a sequence $S$, one solution is to employ the attention mechanism as a discriminator and use $\boldsymbol{e}^{u}$ as a query to assign different weights $\boldsymbol{\alpha}_{i} \in \mathbb{R}^{2}$ to distinct items in $S$. It is worth noting that the first and second dimensions of $\boldsymbol{\alpha}_{i}$ respectively denote the relevance and irrelevance of item $s_{i}$ to user interest $\boldsymbol{e}^{u}$, which can be formulated as follows:
\begin{align}
    \boldsymbol{h}_{i}^{\prime} &= \boldsymbol{h}_{i}\boldsymbol{W}_{3}, \\
    \boldsymbol{\alpha}_{i} &= \sigma\left(\left[\boldsymbol{h}_{i}^{\prime} \| \boldsymbol{e}^{u} \| \boldsymbol{h}_{i}^{\prime} - \boldsymbol{e}^{u} \| \boldsymbol{h}_{i}^{\prime} \odot \boldsymbol{e}^{u}\right]\boldsymbol{W}_{4}\right),
\end{align}
where $\boldsymbol{W}_{3} \in \mathbb{R}^{d \times d}$ and $\boldsymbol{W}_{4} \in \mathbb{R}^{4d \times 2}$ are the trainable parametric matrices, $\odot$ and $\sigma\left(\cdot\right)$ represent the element-wise product and the sigmoid activation function, respectively.

\subsubsection{Soft-level Denoising}
Similar to previous studies \citep{li2020time, xu2019graph}, we introduce a  soft-level denoising module by  carrying out a soft signal denoising layer on the sequence   $S$  based on the attention mechanism, which is followed by a recommender layer. 

\textbf{Weighted with Soft Signal Denoising.} We employ the first dimension of 
$\boldsymbol{\alpha}_{i} \in \mathbb{R}^{2}$ to represent the relevance between the $i$-th item and the user interest, and utilize the normalized weights $\{\alpha_{i}^{0}\}_{i=1}^{n}$ to assign weights to items within the sequence. After that, we obtain the new item representation $\hat{\boldsymbol{h}}_{i} \in \mathbb{R}^{d}$:
\begin{align}
    \alpha_{i}^{0} &= \frac{\exp\left(\alpha_{i}^{0}\right)}{\sum_{j=1}^{n}\exp\left(\alpha_{j}^{0}\right)}, \\
    \hat{\boldsymbol{h}}_{i} &= \alpha_{i}^{0}\boldsymbol{h}_{i}\boldsymbol{W}_{5},
\end{align}
where $\boldsymbol{W}_{5} \in \mathbb{R}^{d \times d}$ is a trainable parametric matrix, and $\hat{\boldsymbol{H}}=\left[\hat{\boldsymbol{h}}_{1}, \hat{\boldsymbol{h}}_{2}, \cdots, \hat{\boldsymbol{h}}_{n}\right] \in \mathbb{R}^{n \times d}$ is the soft denoised sequence representation.

\textbf{Recommender.} 
The recommender serves as a sequence encoder, which can be any mainstream sequence recommendation models. It takes $\hat{\boldsymbol{H}}$ as input and generates the  user representation $\hat{\boldsymbol{e}}^{u}$, which can be formulated as follows:
\begin{align}
    \hat{\boldsymbol{e}}^{u} = \operatorname{F}\left(\hat{\boldsymbol{H}}\right),
\end{align}
where $\operatorname{F}\left(\cdot\right)$ denotes a recommender utilized for sequence representation learning, $e.g.$, BERT4Rec \citep{sun2019bert4rec}, SASRec \citep{kang2018self}, $etc$.

\subsubsection{Hard-level Denoising} We can predict the next item by solely relying on   $\hat{\boldsymbol{e}}^{u}$. However, the influence of noisy items cannot be entirely eliminated  based on the soft-level denoising since it still assigns some weights to them and leading to sub-optimal recommendation performance. To deal with this issue, we propose to further incorporate a hard-level denoising module to improve the learning of user representation. 
To achieve this, we introduce contrastive learning to enhance the robustness of the user interest representation. Nevertheless, the generated hard signals, serving as pseudo-labels, are not fully accurate in detecting noise items due to there is no supervised signals. To address this challenge, we employ the Bayesian Personalized Ranking (BPR) loss \citep{rendle2009bpr,qin2024learning}   to preserve the quality of the generated hard signals.

\textbf{Gumbel-softmax.} Since $\boldsymbol{\alpha}_{i} \in \mathbb{R}^{2}$ serves as a discriminative metric to determine whether an item is noise or not, it can be applied to guide the generation of the hard signals. Due to the need to obtain a binary value and  allow gradient backpropagation,  we opt for the Gumbel-softmax function  to satisfy the requirement, which can be formulated as follows:
\begin{align}
    \Tilde{\boldsymbol{\alpha}}_{i} &= \text{Gumbel-softmax}\left(\boldsymbol{\alpha}_{i}, \tau\right), \\
    \Tilde{\alpha}_{i}^{1} &= \frac{\exp\left(\log\left(\alpha_{i}^{1}\right) + g_{1}\right)/\tau}{\sum_{j=0}^{1}\exp\left(\log\left(\alpha_{i}^{j}\right) + g_{j}\right) / \tau},
\end{align} 
where $g_{j}$ represents a perturbation sampled from the Gumbel distribution, used to enhance the model's robustness. Additionally, $\tau > 0$ serves as a temperature parameter that regulates the selection distribution. When $\tau \rightarrow 0$, $\Tilde{\boldsymbol{\alpha}}_{i}$ approximates a one-hot vector. When $\tau \rightarrow \infty$, $\Tilde{\boldsymbol{\alpha}}_{i}$ approximates a uniform distribution. When $\tau \rightarrow 1$, the Gumbel-softmax function gradually converges to the standard softmax function.
Note that we leverage the second dimension of $\Tilde{\boldsymbol{\alpha}}_{i}=\left[\Tilde{\alpha}_{i}^{0}, \Tilde{\alpha}_{i}^{1}\right] \in \left[0, 1\right]$ as  the hard signal because it denotes the irrelevance between the item representation and the user interest representation. Specifically, if $\Tilde{\alpha}_{i}^{1}=1$, it means that $s_{i}$ is a noisy item, and otherwise $s_{i}$ is a relevant item. 
 
\begin{figure*}[htbp]  
      \centering  
      \includegraphics[width=17cm,keepaspectratio]{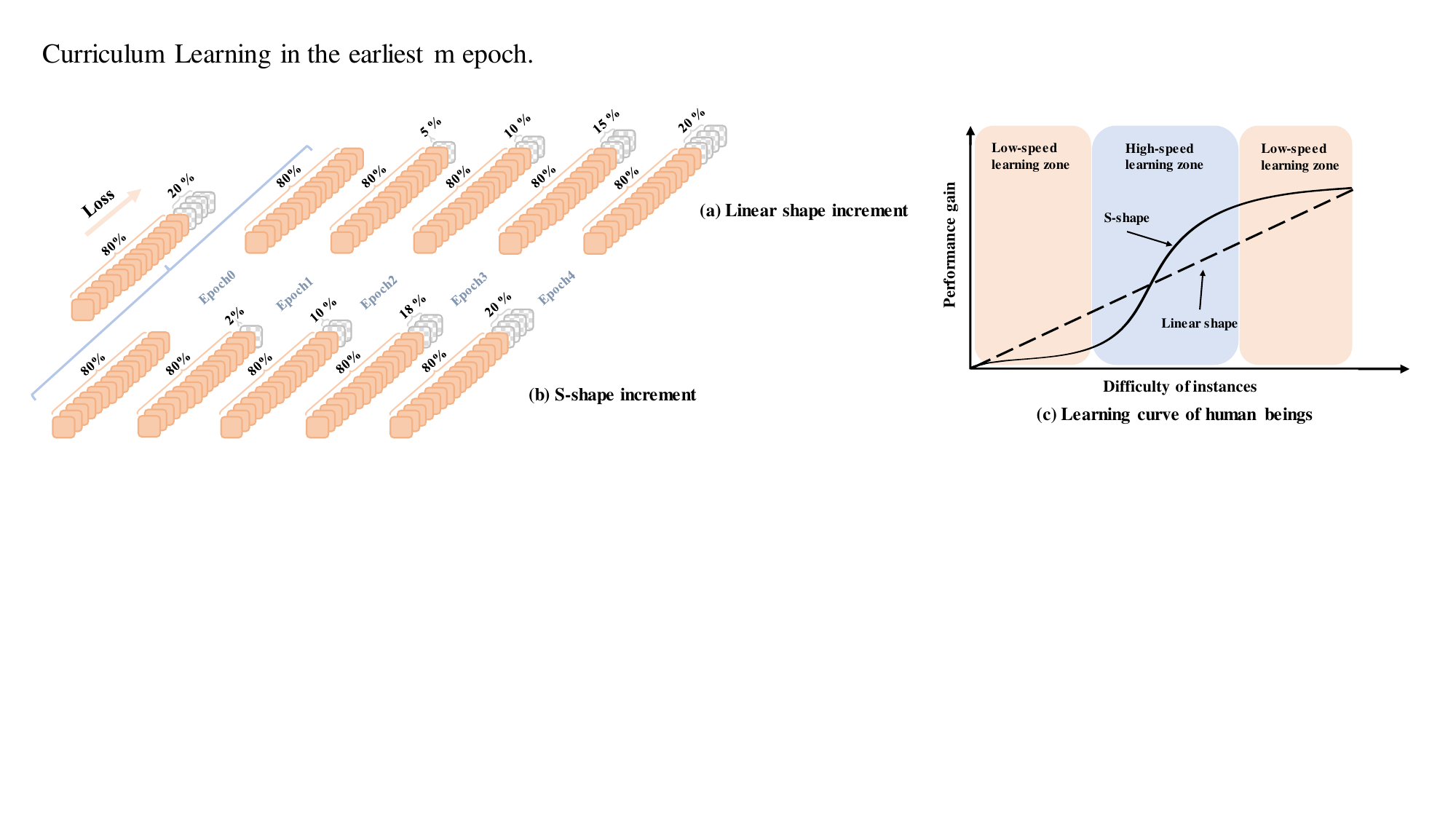} 
      \caption{S-shape curriculum learning vs. Linear shape curriculum learning.} 
      \label{fig:s-shape_curriculum_learning} 
\end{figure*}

\textbf{Dual Loss with Hard Signal Denoising.} In this component, we focus on the task of implementing hard denoising for sequences. It initially uses the generated hard signals $\{\Tilde{\alpha}_{i}^{1}\}_{i=1}^{n}$ to identify noisy items within the sequence. Subsequently, we propose a novel cross-signal contrastive learning to enhance the user interest representation $\hat{\boldsymbol{e}}^{u}$, which is generated via the  soft denoising. In Figure \ref{fig:overall_framework}(c),  $\boldsymbol{h}_{2}$ corresponds to a noisy item, while $\boldsymbol{h}_{1}$ corresponds to a relevant item. The contrastive learning objective is to maximize the correlation between positive sample pairs $\left(\hat{\boldsymbol{e}}^{u}, \boldsymbol{h}_{1}\right)$ and minimize the correlation between negative sample pairs $\left(\hat{\boldsymbol{e}}^{u}, \boldsymbol{h}_{2}\right)$:
\begin{align} \label{eq: SCL}
    \mathcal{L}_{SCL} = -\frac{1}{|S^{+}|}\sum_{s_{i} \in S^{+}}\log\frac{\exp\left(\text{sim}\left(\hat{\boldsymbol{e}^{u}}, \boldsymbol{h}_{i}\right) / \tau\right)}{\sum\limits_{s_{j} \in S}\exp\left(\text{sim}\left(\hat{\boldsymbol{e}^{u}}, \boldsymbol{h}_{j}\right) / \tau\right)}
\end{align}
where $\operatorname{sim}\left(\cdot\right)$ is a cosine similarity function, $\tau$ represents a temperature parameter, and $|S^{+}|$ represents the number of positive samples in the sequence $S$.

However, without supervised signals to ascertain whether items in the sequence are noise, the reliability of the generated hard signals $\{\Tilde{\alpha}_{i}^{1}\}_{i=1}^{n}$ cannot be determined. To tackle this, we use the hard signals to explicit eliminate “noise” items from the original historical interactions, yielding a new sequence representation $\boldsymbol{G}=\left[\boldsymbol{g}_{1}, \boldsymbol{g}_{2}, \cdots, \boldsymbol{g}_{t}\right] \in \mathbb{R}^{t \times d}, t \leq n $. Next, we employ a transformer encoder to encode $\boldsymbol{G}$ in order to acquire the hard denoised user representation $\Bar{\boldsymbol{e}}^{u}$. The process can be formalized as follows:
\begin{align}
    \boldsymbol{G} &= \operatorname{Reorganize}\left(\left(\boldsymbol{1} - \Tilde{\boldsymbol{\alpha}}^{1}\right) \cdot \boldsymbol{H}\right), \\
    \Bar{\boldsymbol{e}}^{u} &= \operatorname{Transformer}\left(\boldsymbol{G}\right),
\end{align}
where $\Tilde{\boldsymbol{\alpha}}^{1}=\left[\Tilde{\alpha}_{1}^{1}, \Tilde{\alpha}_{2}^{1}, \cdots, \Tilde{\alpha}_{n}^{1}\right] \in \mathbb{R}^{d}$ denotes the generated hard denoising signal, and $\operatorname{Reorganize\left(\cdot\right)}$ is a function to reorganize the sequence $S$ after hard denoising. 

After we obtain $\Bar{\boldsymbol{e}}^{u}$, a Bayesian Personalized Ranking (BPR) loss is utilized to maintain the reliability of the generated hard denoised signals:
\begin{align}
    \mathcal{L}_{BPR} = -\frac{1}{\mathcal{|O|}}\sum_{\left(u, i, j\right) \in \mathcal{O}} \ln\sigma\left(\hat{r}_{ui} - \hat{r}_{uj}\right),    \label{eq:BPR loss}
\end{align}
where $\hat{r}_{ui}$ and  $\hat{r}_{uj}$ are the vector inner-product of the positive sample pair $\left(\Bar{\boldsymbol{e}}^{u}, \boldsymbol{h}_{n+1}\right) \in \mathcal{R}^{+}$ and the negative sample pair $\left(\Bar{\boldsymbol{e}}^{u}, \boldsymbol{h}_{j}\right) \in \mathcal{R}^{-}$, respectively. $\mathcal{O}=\{\left(u, i, j\right) | \left(u, i\right) \in \mathcal{R}^{+}, \left(u, j\right) \in \mathcal{R}^{-}\}$.  Since a user's target item is unique, thus the positive sample for the user is $\boldsymbol{h}_{n+1}$, and the randomly sampled negative sample for the user is  $\boldsymbol{h}_{j}$.

\subsection{Prediction}
After applying soft and hard denoising, we obtain a relatively noise-free representation of user interest $\hat{\boldsymbol{e}}^{u}$. Subsequently, within the entire item set space $\mathcal{V}$, we calculate the score between each item $v_{i} \in \mathcal{V}$ and the user's interest representation $\hat{\boldsymbol{e}}^{u}$, denoted as $z_{i}$, to recommend the next item. For each candidate item $v_{i}$, we can use the following formula to quantify its relevance to the sequence:
\begin{align}
    z_{i} = \hat{\boldsymbol{e}}^{u}\boldsymbol{h}_{v_{i}}^\mathsf{T},
\end{align}
then the predicted probability distribution of the next item within the entire item set space $\mathcal{V}$ can be calculated as follows:
\begin{align}
    \hat{y}_{i} = \frac{\exp\left(z_{i}\right)}{\sum\limits_{v_{j} \in \mathcal{V}}\exp\left(z_{j}\right)}.
\end{align}

Subsequently, we define the sequence recommendation task by minimizing the cross-entropy of the predicted results $\hat{y}_{i}$:
\begin{align} \label{eq:CE loss}
    \mathcal{L}_{Rec} = -\sum\limits_{i=1}^{|\mathcal{V}|}\left(y_{i}\log\left(\hat{y}_{i}\right) + \left(1 - y_{i}\right)\log\left(1 - \hat{y}_{i}\right)\right),
\end{align}
where $y_{i}$ denotes the $i$-th one-hot encoding of the ground truth item.

Finally, we use an end-to-end approach to minimize all above loss functions (i.e., Eq.(\ref{eq: SCL}), Eq.(\ref{eq:BPR loss}), Eq.(\ref{eq:CE loss})) to achieve a high performance sequential recommendation model:
\begin{align}\label{eq:loss}
    \text{Loss} = \frac{1}{|\mathcal{U}|}\sum\limits_{u=1}^{|\mathcal{U}|}\left[\mathcal{L}_{Rec} + \lambda\left(\mathcal{L}_{SCL} + \mathcal{L}_{BPR}\right)\right] + \beta\Vert\Theta\Vert_{2}^{2},
\end{align}
where $\lambda$ is a hyperparameter that balances the dual loss  and the recommendation loss,  $\beta$ is a  hyperparameter used to solve the overfitting problem of the model, and $\Theta$ is the total parameter set of the model.

\subsection{S-shape Curriculum Learning}
To boost the performance of our model, we further introduce the curriculum learning \citep{xu2024curriculum,wang2023curriculum,fu2023curriculum}. To be specific, we start to train our model from simple samples and gradually move to more difficult ones. According to Zhang et al. (\citeyear{zhang2022hierarchical}) and Wang et al. (\citeyear{wang2021denoising}), we rank the loss values of all instances in the mini-batch in ascending order. Following the 20/80 principle \citep{sun2020generic}, we divide the mini-batch into ``easy'' instances (i.e., those with lower loss) and ``difficult'' instances  (i.e., those with higher loss), where ``easy'' instances and ``difficult'' instances accounting for 80\% and 20\%, respectively. During the curriculum learning process, the training instances of the model include all the ``easy'' instances together with the $\mu$ percent of ``difficult'' instances, where $\mu$ gradually increases from $0$, progressing with the iterations until epoch $T$ reaches the limit $M$. Both $T$ and $M$ are predefined hyperparameters. 

Existing curriculum learning methods usually assume the training process of increasing ``difficult'' instances in curriculum learning follows a linear pattern. As depicted in Figure \ref{fig:s-shape_curriculum_learning}(a), in these methods, the ``difficult'' instances are fed into model training with a linear shape increment. 
However, different from them, we argue that increasing ``difficult'' instances should follow a S-shape increment, which is inspired by the learning pattern of human beings \citep{murre2014s}.
As shown in Figure \ref{fig:s-shape_curriculum_learning}(c), the learning performance of human beings are lower in the low-speed learning zone, but higher in the high-speed learning zone. Inspired by this phenomenon, we employ fewer ``difficult'' instances in the low-speed learning zones and more ``difficult'' instances in the high-speed learning zone to accommodate different learning stages. To be specific, we utilize a  S-shape function (i.e.,  sigmoid function) to simulate the augmentation process, 
as depicted in Figure \ref{fig:s-shape_curriculum_learning}(b).

\begin{table}[]  
\caption{Statistics of the datasets.}
\scalebox{0.8}{
    \begin{tabular}{lccccc}
    \toprule
    Dataset   & \#Sequence    & \#Users   & \#Items   & \#Avg.length  & \#Sparsity  \\ \midrule
    ML-100k   & 99,287        & 944       & 1,350     & 105.29        & 92.21\%     \\ 
    Beauty    & 198,502       & 22,364    & 12,102    & 8.88          & 99.93\%     \\ 
    Sports    & 296,337       & 35,599    & 18,358    & 8.32          & 99.95\%     \\ 
    Yelp      & 316,354       & 30,432    & 20,034    & 10.40         & 99.95\%     \\ 
    ML-1M     & 999,611       & 6,041     & 3,417     & 165.50        & 95.16\%     \\
    \bottomrule
    \end{tabular}\label{tab:dataset}
}
\end{table}

\section{Experiments}

\subsection{Experimental Settings}
\subsubsection{Datasets}
We conduct extensive experiments on five public benchmark datasets to evaluate the performance of our model. Table \ref{tab:dataset} shows the statistics of the datasets. 
\begin{itemize}
    \item \textbf{MovieLens}\footnote{https://movielens.org/}: MovieLens is a widely used  dataset in the task of sequential recommendation. It contains both user ratings and reviews for movies. In our experiments, we adopt the 100k and 1M versions (named ML-100k and ML-1M, respectively). Note that the average sequence length of this dataset is relatively larger than others. 
    \item \textbf{Amazon-Beauty and Sports}\footnote{http://jmcauley.ucsd.edu/data/amazon}: The dataset contains user historical purchases for a variety of products. Two representative subcategories (i.e., Beauty and Sports) are leveraged in the experiments. Different from MovieLens, users' historical interaction sequences in this dataset is relative short (e.g., 8.88 and 8.32 items on average for Beauty and Sports, respectively).
    \item \textbf{Yelp}\footnote{https://www.yelp.com/dataset}: Yelp is another widely used dataset for sequential recommendation, which records user reviews for various restaurants and bars. As the original dataset is large, we only extract transaction records after 1 January 2019.
\end{itemize}
For all datasets, we keep users' historical interaction sequences in a chronological manner. Following \citep{zhou2022filter, zhang2022hierarchical}, we filter out inactive items with less than 5 interactions and inactive users with less than 5 items. The maximum sequence length for ML-1M is set to 200, and the others are set to 50. Finally, we employ the leave-one-out strategy to split data into training, validation and testing dataset. 

\subsubsection{Evaluation Metrics}
To evaluate the performance of all models, we employ a variety of widely used evaluation metrics \citep{zhang2022hierarchical, zhou2022filter},  including \textit{Hit Ratio} (HR@K), \textit{Normalised Discounted Cumulative Gain} (NDCG@K), and \textit{Mean Reciprocal Rank} (MRR@K). 
To be specific, we utilize HR@\{5, 10, 20\}, NDCG@\{5, 10, 20\} and MRR@\{20\} over the entire item set space to avoid 
the bias introduced by the sampling process \citep{cai2021category, krichene2020sampled}.

\begin{table*}[htbp]
\centering
\caption{Performance comparison of different sequential recommendation methods with (w) or without (w/o) MSDCCL on the five datasets. The best score is in bold. All improvements are statistically significant (i.e., two-sided t-tests with $p < 0.05$).}
\scalebox{0.8}{
    \begin{tabular}{llcccccccccccccc}
    \toprule
    \multirow{2}{*}{Dataset}    &   \multirow{2}{*}{Metric}     &   \multicolumn{2}{c}{GRU4Rec} &  \multicolumn{2}{c}{NARM}    &   \multicolumn{2}{c}{STAMP}   &   \multicolumn{2}{c}{Caser}   &  \multicolumn{2}{c}{SASRec}  &   \multicolumn{2}{c}{CL4SRec}  &   \multicolumn{2}{c}{BERT4Rec}    \\
    \cmidrule{3-16}
    &&   w/o &   w   &   w/o &   w   &   w/o &   w   &   w/o &   w   &   w/o &   w   &   w/o &   w   &   w/o &   w  \\ \midrule
    \multirow{7}{*}{ML-100k}    
                                &   HR@5    &   0.0191  &   \textbf{0.0723} &   0.0180  &   \textbf{0.0756} &   0.0201  &   \textbf{0.0546} &   0.0212  &   \textbf{0.0594} &   0.0191  &   \textbf{0.0682} &   0.0289  &   \textbf{0.0437} &   0.0191  &   \textbf{0.0708} \\
                                &   HR@10   &   0.0286  &   \textbf{0.1249} &   0.0403  &   \textbf{0.1317} &   0.0392  &   \textbf{0.0943} &   0.0339  &   \textbf{0.1051} &   0.0371  &   \textbf{0.1203} &   0.0579  &   \textbf{0.0777} &   0.0414  &   \textbf{0.1270} \\
                                &   HR@20   &   0.0594  &   \textbf{0.2169} &   0.0657  &   \textbf{0.2202} &   0.0700  &   \textbf{0.1702} &   0.0679  &   \textbf{0.1916} &   0.0764  &   \textbf{0.2123} &   0.1107  &   \textbf{0.1404} &   0.0912  &   \textbf{0.2126} \\
                                &   NDCG@5  &   0.0104  &   \textbf{0.0453} &   0.0132  &   \textbf{0.0470} &   0.0115  &   \textbf{0.0331} &   0.0113  &   \textbf{0.0362} &   0.0114  &   \textbf{0.0437} &   0.0174  &   \textbf{0.0267} &   0.0117  &   \textbf{0.0449} \\
                                &   NDCG@10 &   0.0134  &   \textbf{0.0621} &   0.0202  &   \textbf{0.0650} &   0.0176  &   \textbf{0.0459} &   0.0153  &   \textbf{0.0509} &   0.0172  &   \textbf{0.0603} &   0.0265  &   \textbf{0.0378} &   0.0189  &   \textbf{0.0631} \\
                                &   NDCG@20 &   0.0212  &   \textbf{0.0821} &   0.0267  &   \textbf{0.0871} &   0.0253  &   \textbf{0.0648} &   0.0238  &   \textbf{0.0726} &   0.0270  &   \textbf{0.0834} &   0.0398  &   \textbf{0.0536} &   0.0315  &   \textbf{0.0845} \\
                                &   MRR@20  &   0.0109  &   \textbf{0.0494} &   0.0162  &   \textbf{0.0509} &   0.0132  &   \textbf{0.0365} &   0.0119  &   \textbf{0.0409} &   0.0139  &   \textbf{0.0487} &   0.0209  &   \textbf{0.0301} &   0.0157  &   \textbf{0.0498} \\ \midrule
    \multirow{7}{*}{Beauty}    
                                &   HR@5    &   0.0077  &   \textbf{0.0295} &   0.0120  &   \textbf{0.0365} &   0.0080  &   \textbf{0.0421} &   0.0072  &   \textbf{0.0196} &   0.0242  &   \textbf{0.0436} &   0.0208  &   \textbf{0.0249} &   0.0060  &   \textbf{0.0522} \\
                                &   HR@10   &   0.0135  &   \textbf{0.0481} &   0.0209  &   \textbf{0.0565} &   0.0135  &   \textbf{0.0638} &   0.0133  &   \textbf{0.0340} &   0.0386  &   \textbf{0.0632} &   0.0379  &   \textbf{0.0434} &   0.0127  &   \textbf{0.0714} \\
                                &   HR@20   &   0.0256  &   \textbf{0.0753} &   0.0367  &   \textbf{0.0855} &   0.0231  &   \textbf{0.0932} &   0.0235  &   \textbf{0.0550} &   0.0561  &   \textbf{0.0881} &   0.0618  &   \textbf{0.0694} &   0.0204  &   \textbf{0.0955} \\
                                &   NDCG@5  &   0.0045  &   \textbf{0.0194} &   0.0071  &   \textbf{0.0243} &   0.0046  &   \textbf{0.0285} &   0.0044  &   \textbf{0.0120} &   0.0129  &   \textbf{0.0307} &   0.0116  &   \textbf{0.0156} &   0.0037  &   \textbf{0.0378} \\
                                &   NDCG@10 &   0.0064  &   \textbf{0.0254} &   0.0099  &   \textbf{0.0307} &   0.0064  &   \textbf{0.0355} &   0.0064  &   \textbf{0.0166} &   0.0175  &   \textbf{0.0370} &   0.0171  &   \textbf{0.0215} &   0.0059  &   \textbf{0.0439} \\
                                &   NDCG@20 &   0.0094  &   \textbf{0.0322} &   0.0139  &   \textbf{0.0380} &   0.0088  &   \textbf{0.0429} &   0.0090  &   \textbf{0.0219} &   0.0219  &   \textbf{0.0434} &   0.0231  &   \textbf{0.0280} &   0.0078  &   \textbf{0.0500} \\
                                &   MRR@20  &   0.0051  &   \textbf{0.0204} &   0.0077  &   \textbf{0.0249} &   0.0049  &   \textbf{0.0289} &   0.0051  &   \textbf{0.0128} &   0.0122  &   \textbf{0.0307} &   0.0125  &   \textbf{0.0167} &   0.0044  &   \textbf{0.0372} \\ \midrule
    \multirow{7}{*}{Sports}    
                                &   HR@5    &   0.0064  &   \textbf{0.0176} &   0.0099  &   \textbf{0.0199} &   0.0071  &   \textbf{0.0260} &   0.0069  &   \textbf{0.0119} &   0.0113  &   \textbf{0.0258} &   0.0150  &   \textbf{0.0270} &   0.0055  &   \textbf{0.0271} \\
                                &   HR@10   &   0.0114  &   \textbf{0.0293} &   0.0138  &   \textbf{0.0327} &   0.0123  &   \textbf{0.0413} &   0.0115  &   \textbf{0.0199} &   0.0175  &   \textbf{0.0377} &   0.0256  &   \textbf{0.0422} &   0.0104  &   \textbf{0.0387} \\
                                &   HR@20   &   0.0183  &   \textbf{0.0465} &   0.0223  &   \textbf{0.0517} &   0.0182  &   \textbf{0.0629} &   0.0178  &   \textbf{0.0329} &   0.0268  &   \textbf{0.0546} &   0.0418  &   \textbf{0.0638} &   0.0167  &   \textbf{0.0553} \\
                                &   NDCG@5  &   0.0035  &   \textbf{0.0115} &   0.0058  &   \textbf{0.0127} &   0.0046  &   \textbf{0.0169} &   0.0046  &   \textbf{0.0078} &   0.0059  &   \textbf{0.0183} &   0.0088  &   \textbf{0.0177} &   0.0036  &   \textbf{0.0192} \\
                                &   NDCG@10 &   0.0051  &   \textbf{0.0153} &   0.0073  &   \textbf{0.0169} &   0.0062  &   \textbf{0.0218} &   0.0061  &   \textbf{0.0103} &   0.0079  &   \textbf{0.0221} &   0.0122  &   \textbf{0.0226} &   0.0051  &   \textbf{0.0229} \\
                                &   NDCG@20 &   0.0068  &   \textbf{0.0196} &   0.0094  &   \textbf{0.0216} &   0.0077  &   \textbf{0.0272} &   0.0077  &   \textbf{0.0136} &   0.0102  &   \textbf{0.0264} &   0.0163  &   \textbf{0.0280} &   0.0067  &   \textbf{0.0271} \\
                                &   MRR@20  &   0.0036  &   \textbf{0.0122} &   0.0059  &   \textbf{0.0134} &   0.0048  &   \textbf{0.0174} &   0.0049  &   \textbf{0.0083} &   0.0055  &   \textbf{0.0185} &   0.0093  &   \textbf{0.0182} &   0.0040  &   \textbf{0.0192} \\ \midrule
    \multirow{7}{*}{Yelp}    
                                &   HR@5    &   0.0057  &   \textbf{0.0215} &   0.0113  &   \textbf{0.0248} &   0.0060  &   \textbf{0.0234} &   0.0045  &   \textbf{0.0224} &   \textbf{0.0293} &   0.0212  &   0.0276  &   \textbf{0.0307} &   0.0087  &   \textbf{0.0243} \\
                                &   HR@10   &   0.0102  &   \textbf{0.0375} &   0.0187  &   \textbf{0.0416} &   0.0099  &   \textbf{0.0390} &   0.0084  &   \textbf{0.0336} &   0.0352  &   \textbf{0.0357} &   0.0425  &   \textbf{0.0471} &   0.0159  &   \textbf{0.0418} \\
                                &   HR@20   &   0.0184  &   \textbf{0.0631} &   0.0315  &   \textbf{0.0686} &   0.0161  &   \textbf{0.0645} &   0.0146  &   \textbf{0.0517} &   0.0439  &   \textbf{0.0584} &   0.0657  &   \textbf{0.0711} &   0.0273  &   \textbf{0.0695} \\
                                &   NDCG@5  &   0.0034  &   \textbf{0.0133} &   0.0075  &   \textbf{0.0157} &   0.0038  &   \textbf{0.0152} &   0.0028  &   \textbf{0.0158} &   \textbf{0.0251} &   0.0136  &   0.0185  &   \textbf{0.0209} &   0.0054  &   \textbf{0.0151} \\
                                &   NDCG@10 &   0.0048  &   \textbf{0.0184} &   0.0099  &   \textbf{0.0211} &   0.0051  &   \textbf{0.0202} &   0.0040  &   \textbf{0.0194} &   \textbf{0.0270} &   0.0183  &   0.0233  &   \textbf{0.0261} &   0.0077  &   \textbf{0.0207} \\
                                &   NDCG@20 &   0.0068  &   \textbf{0.0248} &   0.0131  &   \textbf{0.0278} &   0.0066  &   \textbf{0.0266} &   0.0055  &   \textbf{0.0239} &   \textbf{0.0292} &   0.0240  &   0.0291  &   \textbf{0.0321} &   0.0105  &   \textbf{0.0277} \\
                                &   MRR@20  &   0.0037  &   \textbf{0.0144} &   0.0081  &   \textbf{0.0167} &   0.0040  &   \textbf{0.0162} &   0.0031  &   \textbf{0.0163} &   \textbf{0.0250}  &   0.0146 &   0.0191  &   \textbf{0.0214} &   0.0060  &   \textbf{0.0163} \\ \midrule
    \multirow{7}{*}{ML-1M}    
                                &   HR@5    &   0.0194  &   \textbf{0.1689} &   0.0151  &   \textbf{0.1676} &   0.0232  &   \textbf{0.1464} &   0.0104  &   \textbf{0.1394} &   0.0397  &   \textbf{0.1455} &   0.0369  &   \textbf{0.0754} &   0.0224  &   \textbf{0.1351} \\
                                &   HR@10   &   0.0373  &   \textbf{0.2473} &   0.0349  &   \textbf{0.2457} &   0.0440  &   \textbf{0.2145} &   0.0215  &   \textbf{0.2158} &   0.0666  &   \textbf{0.2251} &   0.0684  &   \textbf{0.1294} &   0.0495  &   \textbf{0.2095} \\
                                &   HR@20   &   0.0690  &   \textbf{0.3445} &   0.0591  &   \textbf{0.3430} &   0.0677  &   \textbf{0.3010} &   0.0589  &   \textbf{0.3173} &   0.1007  &   \textbf{0.3260} &   0.1204  &   \textbf{0.2064} &   0.0980  &   \textbf{0.3062} \\
                                &   NDCG@5  &   0.0135  &   \textbf{0.1139} &   0.0080  &   \textbf{0.1124} &   0.0150  &   \textbf{0.0993} &   0.0063  &   \textbf{0.0899} &   0.0207  &   \textbf{0.0936} &   0.0216  &   \textbf{0.0471} &   0.0132  &   \textbf{0.0873} \\
                                &   NDCG@10 &   0.0190  &   \textbf{0.1392} &   0.0144  &   \textbf{0.1376} &   0.0218  &   \textbf{0.1212} &   0.0099  &   \textbf{0.1145} &   0.0294  &   \textbf{0.1192} &   0.0317  &   \textbf{0.0644} &   0.0218  &   \textbf{0.1112} \\
                                &   NDCG@20 &   0.0270  &   \textbf{0.1636} &   0.0205  &   \textbf{0.1621} &   0.0278  &   \textbf{0.1430} &   0.0194  &   \textbf{0.1401} &   0.0379  &   \textbf{0.1447} &   0.0447  &   \textbf{0.0837} &   0.0339  &   \textbf{0.1356} \\
                                &   MRR@20  &   0.0159  &   \textbf{0.1129} &   0.0100  &   \textbf{0.1113} &   0.0168  &   \textbf{0.0988} &   0.0091  &   \textbf{0.0908} &   0.0203  &   \textbf{0.0940} &   0.0243  &   \textbf{0.0501} &   0.0169  &   \textbf{0.0881} \\
    \bottomrule
    \label{tab:overall_mainstream_model}
    \end{tabular}
}
\end{table*}
 
\subsubsection{Competing Models}
To demonstrate the effectiveness of MSDCCL, we fuse it with six base models of sequential recommendation, and compare them with their counterparts. These base models are given as follows: 
\begin{itemize}
    \item \textbf{GRU4Rec} \citep{hidasi2015session}. This method applies the gated recurrent unit (GRU) to model sequential data for recommendation. It modified the original GRU by introducing a tailored ranking loss function and session-parallel mini-batches.  
    \item \textbf{NARM} \citep{li2017neural}. It utilizes a hybrid encoder to capture the sequential behavior and the main purpose of a user in the current sequence. In addition, a bi-linear matching module is introduced to obtain recommendation scores.    
    \item \textbf{STAMP} \citep{liu2018stamp}. This method proposes to capture both users' general interests and current interests. It leverages the long-term memory of the whole session as the general interests, and the short-term memory of the last-clicks as the current interests.      
    \item \textbf{CASER} \citep{tang2018personalized}. It employs convolutional operation to capture sequential patterns and models users' recent items from both time and latent dimensions.  
    \item \textbf{SASRec} \citep{kang2018self}. It proposes a self-attention based sequential recommendation method, which utilizes self-attention mechanism  to simultaneously capture both long-term semantics as well as a few important recent items. 
    \item \textbf{CL4SRec} \citep{xie2022contrastive}. It is a contrastive learning based sequential recommender which attempts to learn better sequence representations by leveraging the contrastive learning framework to obtain self-supervision signals.
    \item \textbf{BERT4Rec} \citep{sun2019bert4rec}. This method applies the deep bidirectional self-attention to model user behavior sequence, and utilizes the Cloze task as the objective in sequential recommendation.

\end{itemize}

In addition, we also compare the performance of our fused model (i.e., MSDCCL+BEAT4Rec) with the following state-of-the-art denoising models, including:
\begin{itemize}
    \item \textbf{DSAN} \citep{yuan2021dual}. It learns target embedding by exploring item-level interaction and correlation within a session. In addition, an adaptively sparse attention is used to  mitigate the influence of irrelevant items.  
    \item \textbf{FMLP-Rec} \citep{zhou2022filter}. To obtain  more robust sequence representations, it utilizes the fast fourier transform (FFT) \citep{ricardo2018scheme} and its inverse transform process to attenuate the effect of noise within a sequence.
    \item \textbf{HSD+BERT4Rec} \citep{zhang2022hierarchical}. It removes the noisy items in the original sequence by generating sequence-level and user-level signals, and learns better sequence representations by using the reassembled sequence.
    \item \textbf{AC-BERT4Rec} \citep{zhou2023attention}. This model introduces a novel spatial calibrator and an adversarial calibrator  to capture spatial relationships between items and redistribute attention weights based on the contributions of different items to predict the next item. In this model, we take BEAT4Rec as the backbone.
\end{itemize}

\begin{table*}[htbp]
\centering
\caption{Performance comparison between MSDCCL with the best  base model (i.e., BERT4Rec) and the state-of-the-art denoising methods on the five datasets. The best score and the second best score in each column are in bold and underlined, respectively. Here * denotes statistically significant improvement (measured by a two-sided t-test with $p < 0.05$) over the best baseline. OOM: Out of memory on 24GB DGX.}
\begin{tabular}{llccccccccc}
    \toprule
    Dataset &   \multicolumn{3}{l}{Model}   &   HR@5    &   HR@10   &   HR@20   &   NDCG@5  &   NDCG@10 &   NDCG@20 &   MRR@20  \\ \midrule
    \multirow{4}{*}{ML-100k}    
        &   \multicolumn{3}{l}{DSAN (AAAI'21)}           &   0.0201  &   0.0435   &   0.0700  &   0.0115  &   0.0188  &   0.0254  &   0.0133   \\
        &   \multicolumn{3}{l}{FMLP-Rec (WWW'22)}        &   0.0170  &   0.0477   &   0.0764  &   0.0117  &   0.0216  &   0.0288  &   0.0160   \\
        &   \multicolumn{3}{l}{HSD+BERT4Rec (CIKM'22)}   &   \underline{0.0339}  &   \underline{0.0732}  &   \underline{0.1294}  &   \underline{0.0178}  &   \underline{0.0305}  &   \underline{0.0447}  &   \underline{0.0218}   \\
        &   \multicolumn{3}{l}{AC-BERT4Rec (CIKM'23)}   &   0.0212  &   0.0445  &   0.0877  &   0.0113  &   0.0188  &   0.0320  &   0.0147   \\
        &   \multicolumn{3}{l}{MSDCCL+BERT4Rec}           &   \textbf{0.0708}*     &   \textbf{0.1270}* &   \textbf{0.2126}* &   \textbf{0.0449}* &   \textbf{0.0631}* &   \textbf{0.0845}* &   \textbf{0.0498}* \\ \midrule
    \multirow{4}{*}{Beauty}    
        &   \multicolumn{3}{l}{DSAN (AAAI'21)}           &   0.0092  &   0.0152   &   0.0264  &   0.0058  &   0.0077  &   0.0105  &   0.0062   \\
        &   \multicolumn{3}{l}{FMLP-Rec (WWW'22)}        &   0.0095  &   0.0166   &   0.0284  &   0.0056  &   0.0078  &   0.0107  &   0.0060   \\
        &   \multicolumn{3}{l}{HSD+BERT4Rec (CIKM'22)}   &   \underline{0.0261}  &   \underline{0.0447}  &   \underline{0.0683}  &   \underline{0.0147}  &   \underline{0.0207}  &   \underline{0.0266}  &   \underline{0.0151}   \\
        &   \multicolumn{3}{l}{AC-BERT4Rec (CIKM'23)}   &   0.0200  &   0.0371  &   0.0609  &   0.0120  &   0.0175  &   0.0235  &   0.0133   \\
        &   \multicolumn{3}{l}{MSDCCL+BERT4Rec}           &   \textbf{0.0522}*     &   \textbf{0.0714}* &   \textbf{0.0955}* &   \textbf{0.0378}* &   \textbf{0.0439}* &   \textbf{0.0500}* &   \textbf{0.0372}* \\ \midrule
    \multirow{4}{*}{Sports}    
        &   \multicolumn{3}{l}{DSAN (AAAI'21)}           &   0.0061  &   0.0105   &   0.0215  &   0.0042  &   0.0056  &   0.0084  &   0.0049   \\
        &   \multicolumn{3}{l}{FMLP-Rec (WWW'22)}        &   0.0068  &   0.0117   &   0.0180  &   0.0044  &   0.0059  &   0.0075  &   0.0046   \\
        &   \multicolumn{3}{l}{HSD+BERT4Rec (CIKM'22)}   &   \underline{0.0120}  &   0.0190  &   0.0303  &   \underline{0.0078}  &   \underline{0.0100}  &   0.0129  &   \underline{0.0081}   \\
        &   \multicolumn{3}{l}{AC-BERT4Rec (CIKM'23)}   &   0.0112  &   \underline{0.0203}  &   \underline{0.0351}  &   0.0069  &   0.0099  &   \underline{0.0136}  &   0.0077   \\
        &   \multicolumn{3}{l}{MSDCCL+BERT4Rec}           &   \textbf{0.0271}*     &   \textbf{0.0387}* &   \textbf{0.0553}* &   \textbf{0.0192}* &   \textbf{0.0229}* &   \textbf{0.0271}* &   \textbf{0.0192}* \\ \midrule
    \multirow{4}{*}{Yelp}    
        &   \multicolumn{3}{l}{DSAN (AAAI'21)}           &   0.0269  &   0.0369   &   0.0541  &   \underline{0.0211}  &   0.0242  &   0.0285  &   \underline{0.0216}   \\
        &   \multicolumn{3}{l}{FMLP-Rec (WWW'22)}        &   0.0203  &   0.0294   &   0.0436  &   0.0142  &   0.0171  &   0.0207  &   0.0144   \\
        &   \multicolumn{3}{l}{HSD+BERT4Rec (CIKM'22)}   &   \textbf{0.0292}*  &   0.0408  &   0.0593  &   \textbf{0.0223}*  &   \textbf{0.0260}*  &   \underline{0.0307}  &   \textbf{0.0228}*   \\
        &   \multicolumn{3}{l}{AC-BERT4Rec (CIKM'23)}   &   \underline{0.0286}  &   \textbf{0.0445}*  &   \textbf{0.0700}*  &   0.0194  &   \underline{0.0245}  &   \textbf{0.0309}*  &   0.0202   \\
        &   \multicolumn{3}{l}{MSDCCL+BERT4Rec}           &   0.0243     &   \underline{0.0418} &   \underline{0.0695} &   0.0151 &   0.0207 &   0.0277 &   0.0163 \\ \midrule
    \multirow{4}{*}{ML-1M}    
        &   \multicolumn{3}{l}{DSAN (AAAI'21)}           &   0.0098  &   0.0336   &   0.0651  &   0.0048  &   0.0122  &   0.0200  &   0.0081   \\
        &   \multicolumn{3}{l}{FMLP-Rec (WWW'22)}        &   0.0210  &   0.0449   &   0.0707  &   0.0120  &   0.0199  &   0.0263  &   0.0142   \\
        &   \multicolumn{3}{l}{HSD+BERT4Rec (CIKM'22)}   &   \underline{0.0477}  &   \underline{0.0886}  &   \underline{0.1399}  &   \underline{0.0297}  &   \underline{0.0429}  &   \underline{0.0558}  &   \underline{0.0328}   \\
        &   \multicolumn{3}{l}{AC-BERT4Rec (CIKM'23)}   &   OOM  &   OOM  &   OOM  &   OOM  &   OOM  &   OOM  &   OOM   \\
        &   \multicolumn{3}{l}{MSDCCL+BERT4Rec}           &   \textbf{0.1351}*     &   \textbf{0.2095}* &   \textbf{0.3062}* &   \textbf{0.0873}* &   \textbf{0.1112}* &   \textbf{0.1356}* &   \textbf{0.0881}* \\ 
    \bottomrule
    \label{tab:overall_denoising_model}
    \end{tabular}
\end{table*}

\subsubsection{Implementation Details}
Similar to previous models \citep{yuan2021dual, zhang2022hierarchical}, we set the embedding and mini-batch sizes for all models to 100 and 256, respectively. The learning rate of Adam optimizer is set to $10^{-3}$. We tune the regularization hyperparameter $\beta$ in $\{0, 10^{-3}, 10^{-4}\}$ on the validation set.
The initial temperature parameter $\tau$ in the cross-signal contrastive learning module  is set to 0.5, which is annealed after every 40 batches.  All embedding parameters are initialized with a Gaussian distribution.
For all baseline methods, we either report their performances in the original papers or optimally tuned on the validation data.
The early-stopping training strategy is employed to prevent model overfitting, which stops training when the HR@20 metric decreases continuously over 10 epochs on the validation dataset.


\subsection{Overall Performance Comparison}
The results of different models on datasets are shown in Table \ref{tab:overall_mainstream_model} and Table \ref{tab:overall_denoising_model}. The best results are in bold and the second best results are underlined. We conduct the two-sided t-test with $p < 0.05$ to verify the statistical significance for the  performance improvement. 
From the experimental results, we draw the following conclusions:
\begin{itemize}
    \item Table \ref{tab:overall_mainstream_model} shows the performance of different sequential recommendation methods with or without our proposed MSDCCL approach over all datasets. We observe that the base models fused with MSDCCL yield significant improvements in most cases when comparing with their counterparts.     
    The results verify the effectiveness of our proposed method which can extract better user preferences together with alleviating the noise issue. To be specific, the relative improvements of MSDCCL+BERT4Rec  over BERT4Rec in terms of the metrics HR@20, NDCG@20, MRR@20 are 133.11\%, 168.25\%, 217.20\% respectively on the dataset ML-100k. Similar relative improvements can also be observed on other datasets.
    
     \item Table \ref{tab:overall_denoising_model} reports the performance comparison of all denoising approaches on all five datasets. We can see that on the ML-100k, Beauty, Sports and ML-1M datasets, the performance of MSDCCL+BERT4Rec, which employs both the soft and hard denoising strategy, consistently outperforms other denoising models relying solely on the soft denoising strategy (e.g., DSAN, FMLP-Rec and AC-BERT4Rec) or hard denoising strategy (e.g., HSD+BEART4Rec). 
     It is worth noting that the performance of MSDCCL+BERT4Rec is suboptimal on the Yelp dataset. This may be due to the fact that the denoising signals produced by MSDCCL are more difficult to fuse with BERT4Rec than the denoising signals produced by other models on certain datasets. In summary, these experimental results demonstrate the superiority of our denoising model over other sequence denoising recommendation models in most cases.


\end{itemize}

\begin{table*}[htb]
\centering
\caption{Ablation study of our model, we report HR@20, NDCG@20 and MRR@20 on five datasets. The best results are highlighted in bold, and the second best results are underlined.}
\scalebox{0.9 }{
    \begin{tabular}{llccccc}
    \toprule
    Dataset &   Metric    &   Full   &   w/o DL   &   w/o TS  &   w/o BPR &   w/o CL  \\ \midrule
    \multirow{3}{*}{ML-100k}
        &   HR@20   &   \textbf{0.2126} &   0.1783  &   0.2054  &   0.1991  &   \underline{0.2079} \\
        &   NDCG@20 &   \textbf{0.0845} &   0.0678  &   0.0823  &   0.0793  &   \underline{0.0826} \\
        &   MRR@20  &   \textbf{0.0498} &   0.0380  &   \underline{0.0489}  &   0.0468  & 0.0486 \\  \midrule
    \multirow{3}{*}{Beauty}
        &   HR@20   &   \underline{0.0955} &   0.0639  &   \textbf{0.0957}  &   0.0937  &   0.0945 \\
        &   NDCG@20 &   \textbf{0.0500} &   0.0279  &   \textbf{0.0500}  &   0.0491  &   \underline{0.0492} \\
        &   MRR@20   &   \textbf{0.0372} &   0.0180  &   \underline{0.0371}  &   0.0365  & 0.0364 \\  \midrule
    \multirow{3}{*}{Sports}
        &   HR@20   &   \textbf{0.0553} &   0.0461  &   0.0548  &   \underline{0.0550} &   \underline{0.0550} \\
        &   NDCG@20 &   \textbf{0.0271} &   0.0193  &   0.0264  &   \underline{0.0268}  &   0.0267 \\
        &   MRR@20   &   \textbf{0.0192} &   0.0120  &   0.0185  &   \underline{0.0191}  & 0.0188 \\  \midrule
    \multirow{3}{*}{Yelp}
        &   HR@20   &   \textbf{0.0695} &   \underline{0.0689}  &   0.0687  &   \underline{0.0689}  &   0.0671 \\
        &   NDCG@20 &   \textbf{0.0277} &   \underline{0.0276}  &   0.0274  &   0.0273  &   0.0267 \\
        &   MRR@20   &   \textbf{0.0163} &   \textbf{0.0163}  &   \underline{0.0162}  &   0.0160  & 0.0157 \\  \midrule
    \multirow{3}{*}{ML-1M}
        &   HR@20   &   \textbf{0.3062} &   0.2826  &   0.3033  &   \underline{0.3034}  &   0.3031 \\
        &   NDCG@20 &   \textbf{0.1356} &   0.1249  &   0.1337  &   \underline{0.1346}  &   0.1336 \\
        &   MRR@20   &   \textbf{0.0881} &   0.0811  &   0.0866  &   \underline{0.0877}  & 0.0864 \\  
    \bottomrule
    \label{tab:ablation}
    \end{tabular}}
\end{table*}

\subsection{Ablation Study}
To investigate the role of each component in our proposed MSDCCL, we perform an ablation study by removing each component from the entire model for comparison.  Note that we utilize BERT4Rec as the backbone for all variants.
To be specific, we consider the following variants of our method: 
\begin{itemize}    
    \item  \textbf{w/o DL}: It discards the Dual Loss with Hard Signal Denoising module, which is developed to explicitly eliminate noisy items from the original sequence with hard signals.  
    \item  \textbf{w/o TS}: It denotes the variant which drops the target signal from the Target-Aware User Interest Extractor module.
    \item  \textbf{w/o BPR}:  This variant ignores the supervised loss (i.e., BPR loss) in the Dual Loss with Hard Signal Denoising module.
    \item   \textbf{w/o CL}: It removes the curriculum learning module from our model, which indicates that the model will be trained without considering the difficulty of samples. 
    \item \textbf{Full}: This model indicates our full model without dropping any component.
\end{itemize}
 
As shown in Table \ref{tab:ablation}, we observe that discarding each component of our model MSDCCL will lead to a performance drop in most cases. Specifically, removing the Dual Loss with Hard Signal Denoising module (i.e., w/o DL) will cause a considerable performance decline on all datasets. 
Compared to MSDCCL, the relative performance decrements of w/o DL are 16.13\%, 19.76\%, 23.69\% in terms of the metrics HR@20, NDCG@20, MRR@20 on the dataset ML-100k. Similar results can be observed on the other datasets. This reveals that the Dual Loss with Hard Signal Denoising module is indeed critical for alleviating the impact of noisy items. 
Besides, removing the target signal from the Target-Aware User Interest Extractor module (i.e., w/o TS) will also result in an inferior performance, which demonstrates the usefulness of taking the target signal for learning better representations of users. 
It is interesting to notice that when we remove the supervised loss in the Dual Loss with Hard Signal Denoising module (i.e., w/o BPR), the drop of the performance is less than that of w/o DL. This indicates that both unsupervised loss (i.e., cross-signal contrastive learning loss) and supervised loss (i.e., BPR loss) are necessary in our model. 
At last, if we exclude the curriculum learning module, the performance will decrease consistently, which indicates the importance of incorporating the extended curriculum learning  in our proposed model.

\begin{figure*}[htbp]  
      \centering  
      \includegraphics[width=17cm,keepaspectratio]{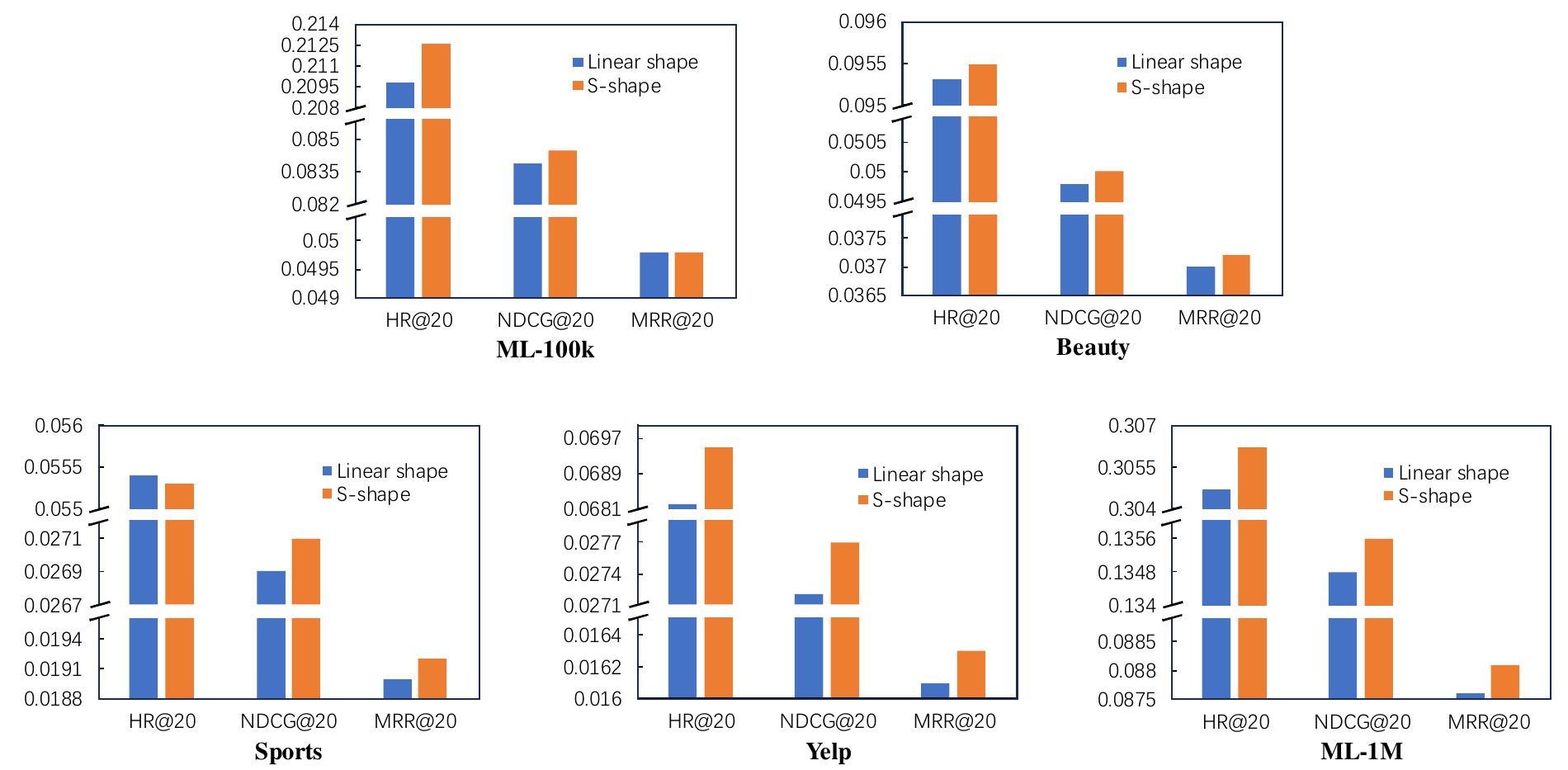} 
      \caption{Analysis on the effectiveness of different increment patterns in curriculum learning, specifically linear shape versus S-shape.} 
      \label{fig:linear_or_s-shape_curve} 
\end{figure*}

\subsection{Effect of S-shape Curriculum Learning}
In this section, we further investigate the effectiveness of developing the S-shape  curriculum learning. To the end, we compare the performance of two different increment strategies of training samples (i.e., linear shape increment and S-shape increment) during the curriculum learning process. From Figure \ref{fig:linear_or_s-shape_curve}, we can see that  MSDCCL equipped with S-shape increment performs better than its counterpart equipped with linear increment. This indicates that aligning the increment of training samples with the learning pattern of human beings  can provide better learning capacity  of curriculum learning in sequential recommendation. 

\subsection{Comparison with Different Sequence Length}


\begin{figure*}[htb!]  
      \centering  
      \includegraphics[width=17cm,keepaspectratio]{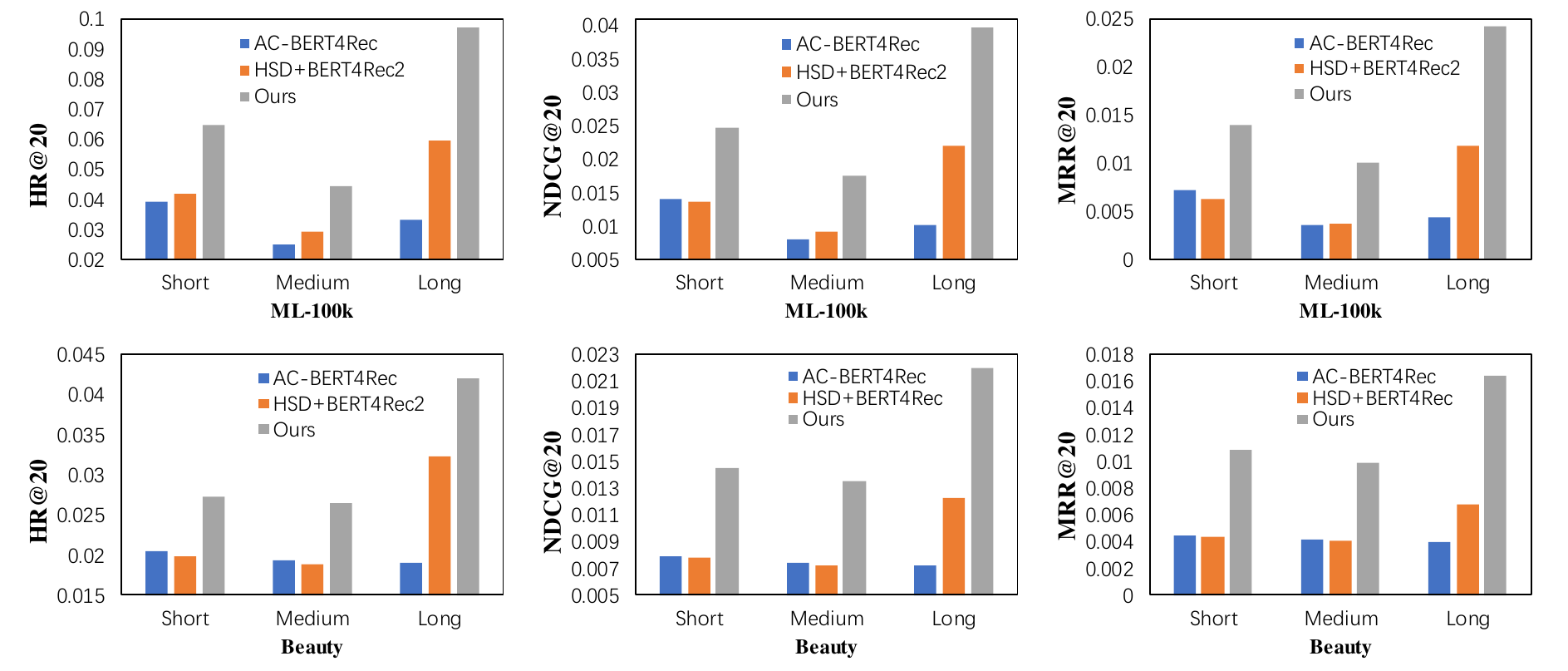} 
      \caption{Performance of our model and the two best performing baselines (i.e., AC-BERT4Rec, HSD+BERT4Rec) under different sequence lengths on  ML-100k and Beauty.} 
      \label{fig:different_seq_length} 
\end{figure*}

The sequence length reflects the sparsity of user interaction behaviors. 
To evaluate the performance of our model trained with different sequence lengths, we evenly split sequences of a dataset (e.g., ML-100k and Beauty)  into three groups (i.e., short, medium and long) based on their lengths. We compare the performance of our model with the two best performing baselines, including AC-BERT4Rec and HSD+BERT4Rec. 
According to the results demonstrated in Figure \ref{fig:different_seq_length}, we see that AC-BERT4Rec obtains superior performance on short sequences as compared with that on both medium and long sequences. This is because these medium and long sequences would contain more noisy items, and AC-BERT4Rec shows inferior results due to the assignment of attention weights to them. On the contrary, the performance of HSD+BERT4Rec on long sequences is much higher than that on both short and medium sequences. This is due to that HSD+BERT4Rec takes a hard denoising strategy which directly discards noisy items in  sequences. Since there is no supervised signal for sequence denoising, this method would inevitably overlook relevant items and the problem becomes more serious when the sequence is short. 
Our model is consistently superior to both AC-BERT4Rec and HSD+BERT4Rec on all three groups. 
The main reason is that our model can make full use of advantages of both soft and hard denoising strategies where each of them is leveraged to guide the learning process of each other.   


\subsection{Impact of Training Set Proportion}

\begin{figure*}[htb]  
      \centering    \includegraphics[width=17cm,keepaspectratio]{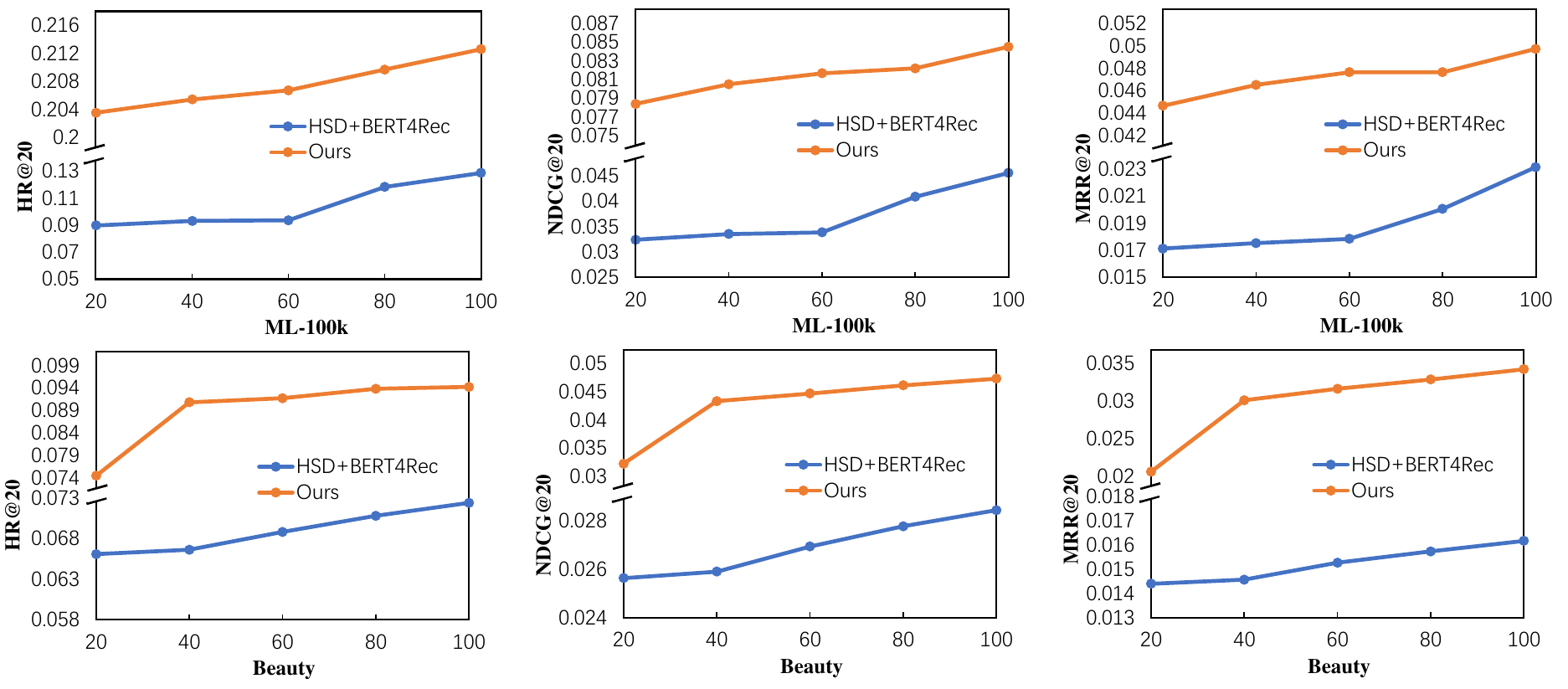}  
      \caption{Performance of our model and the best performing baseline HSD+BERT4Rec with different training set proportion on  ML-100k and Beauty.}  \label{fig:different_train_set_ratio}  
\end{figure*}

To investigate the effectiveness of our proposed model with different training set ratios, we compare the performance of our model with the best performing baseline HSD+BERT4Rec on the dataset ML-100k and Beauty. Figure \ref{fig:different_train_set_ratio} shows the results when we vary the training set proportion from 20\% to 100\% with a step size of 20\%. It can be seen from the results that the performance of both methods increase gradually with the growth of training data. In addition, our model significantly  outperforms HSD+BERT4Rec  under all training set ratios, which demonstrates the effectiveness of our model in real-world applications.

\begin{figure*}[htbp]  
      \centering  
      \includegraphics[width=18cm,keepaspectratio]{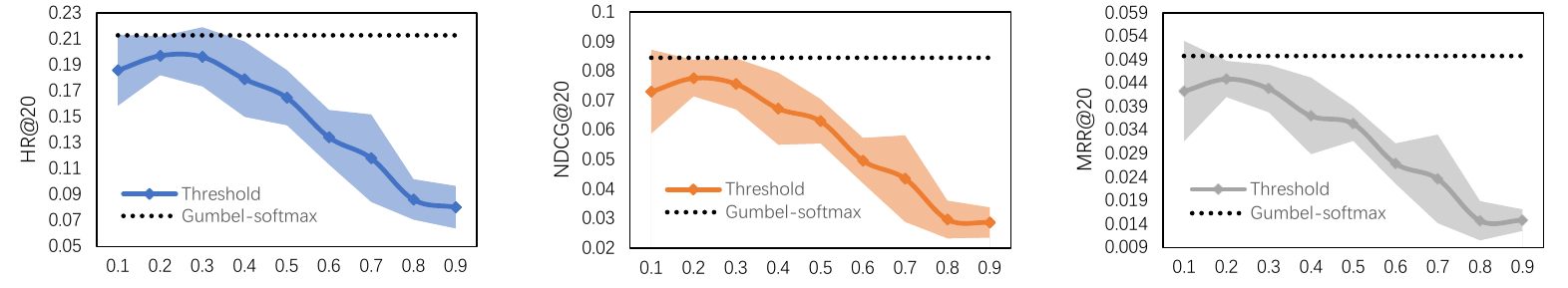} 
      \caption{Performance Comparison of our model with the Gumbel-softmax strategy and the clipping strategy on the ML-100k dataset.}
      \label{fig:gumbel_threshold} 
\end{figure*}
\subsection{Impact of the Gumbel-softmax Strategy}
To validate the effectiveness of incorporating the Gumbel-softmax strategy in our proposed method, we replace it  with a simple clipping strategy. 
We vary the value of the clipping threshold  from 0.1 to 0.9 with a step size of 0.1, and the performance on the ML-100k dataset is demonstrated in Figure \ref{fig:gumbel_threshold}. We can observe that the model performance with  the clipping strategy improves if a higher threshold value is used and reaches a peak when  the threshold value is 0.2. Then the performance starts to drop continuously as the threshold value increases. In addition, we can clearly observe that the model performance with the Gumbel-softmax strategy is consistently superior to that of the clipping strategy, which suggests that it is beneficial to employ the Gumbel-softmax strategy in our proposed method. 
This is probably because the clipping strategy always obtain the same output while the Gumbel-softmax strategy might get probability output due to the incorporation of the Gumbel noise.

\begin{figure}[htbp]  
      \centering  
      \includegraphics[width=8.5cm,keepaspectratio]{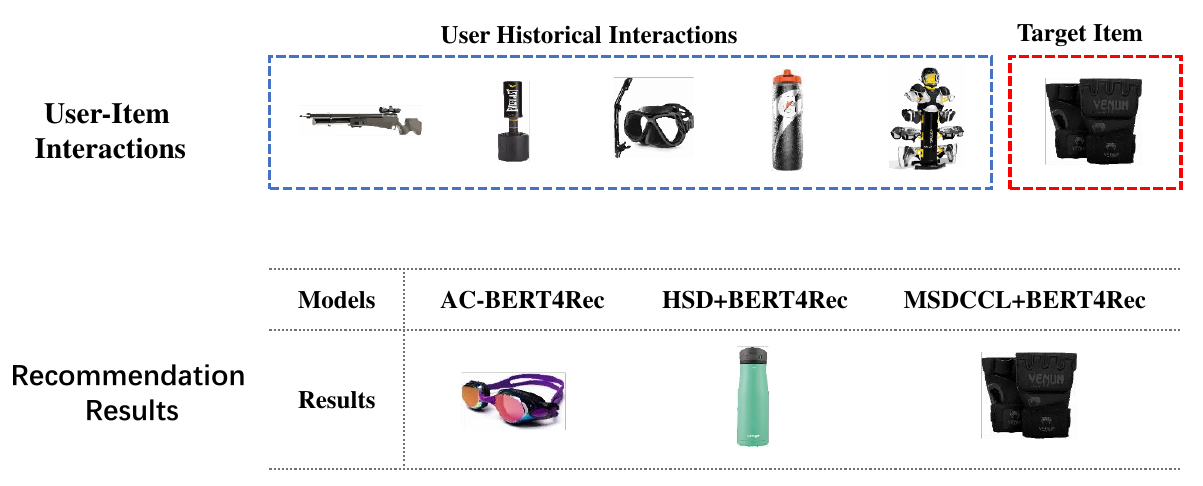} 
      \caption{A case study on Sports to demonstrate the effectiveness of our model.} 
      \label{fig:case_study} 
\end{figure}
\subsection{Case Study}
We further conduct a case study to investigate the advantages of MSDCCL in comparison other denoising models (i.e., soft and hard denoising methods). Figure \ref{fig:case_study} shows a historical interaction sequence (in the blue dotted box) and the target item (in the red dotted box) of  an indoor sport enthusiast. Among all methods, only the proposed method makes the correct recommendation. The main reason is that solely employing the soft denoising strategy (e.g., AC-BERT4Rec) would still assign  low attention weights to noise items such as  ``gun'' and thus affect the recommending performance. In contrast, utilizing the hard denoising strategy (e.g., HSD+BERT4Rec) would discard informative information due to the lack of supervised signal for sequence denoising. Different from these methods, our model can comprehensively leverage the advantages of both soft and hard denoising strategies, thus effectively boosting the capability of sequence denoising and accurately capturing user intent for recommendation. 

\subsection{Hyperparameter Sensitivity}

\begin{figure*}[htbp]  
      \centering  
      \includegraphics[width=17cm,keepaspectratio]{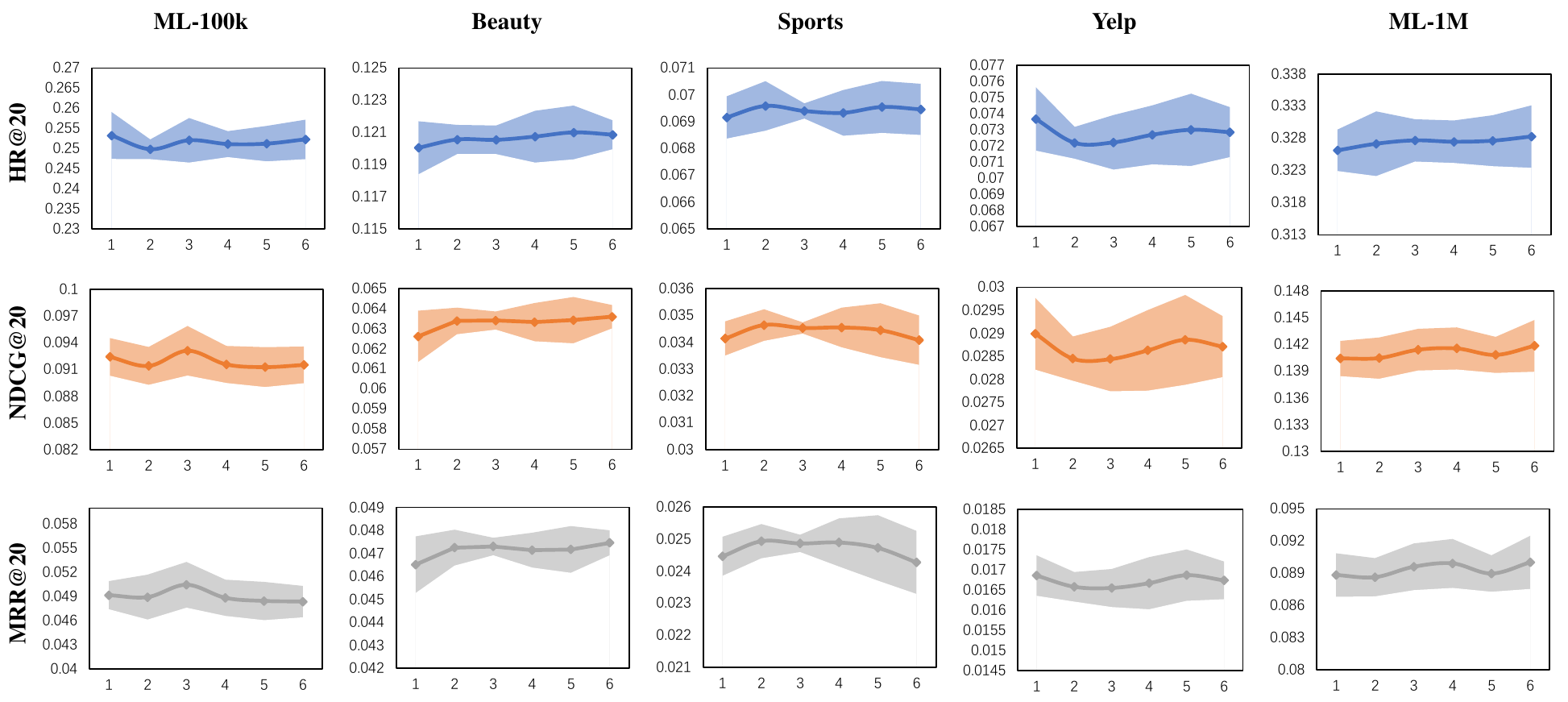} 
      \caption{Performance (HR@20, NDCG@20 and MRR@20) comparison w.r.t hyperparameters $m$ on five datasets.} 
      \label{fig:hyperparameter_sensitivity_m} 
\end{figure*}
\begin{figure*}[htbp]  
      \centering  
      \includegraphics[width=17cm,keepaspectratio]{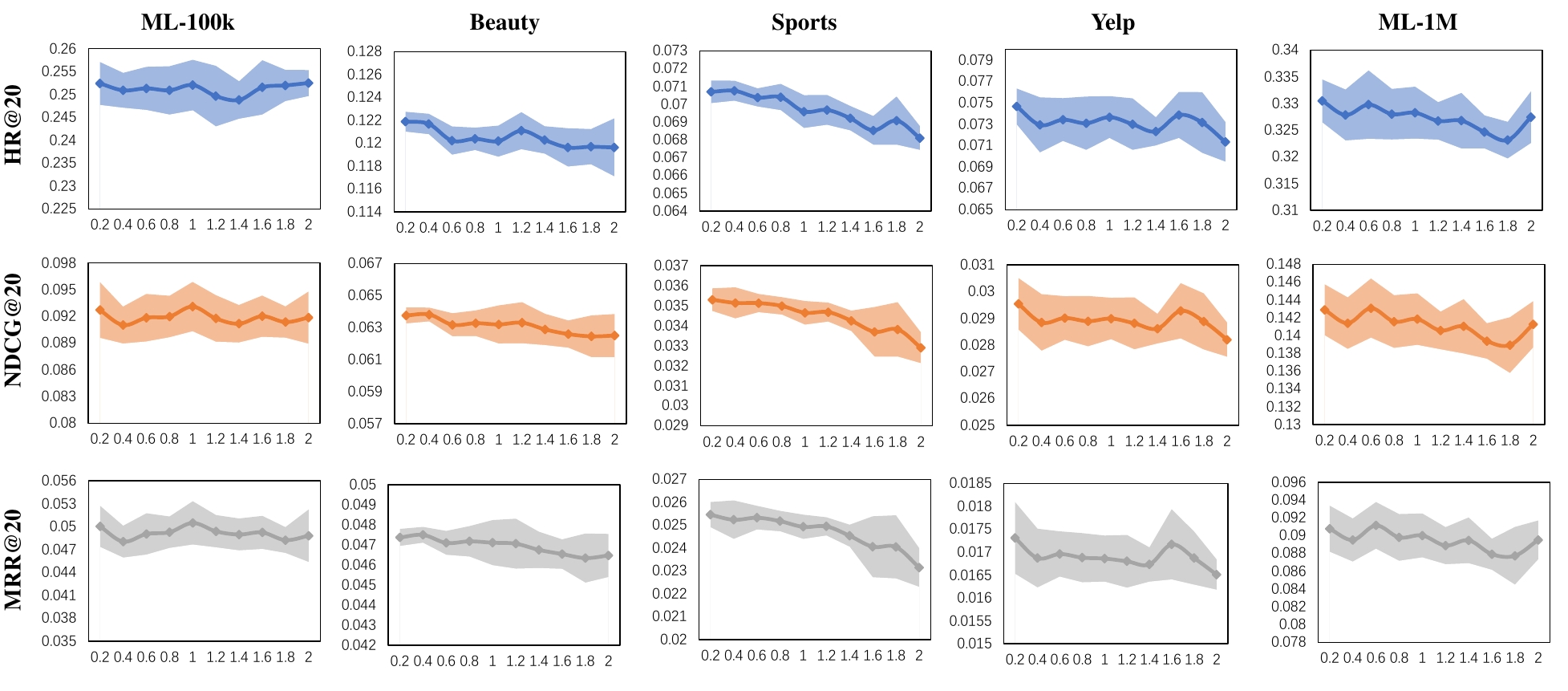} 
      \caption{Performance (HR@20, NDCG@20 and MRR@20) comparison w.r.t hyperparameters $\lambda$ on five datasets.} 
      \label{fig:hyperparameter_sensitivity_lambda} 
\end{figure*}

In this section, we investigate the impact of hyperparameters in MSDCCL. We first explore the effect of $m$, which plays a critical role in modeling short user interests. Then, we analyze the impact of $\lambda$ which is introduced to balance  the importance between the dual loss with hard signal denoising and the recommendation loss. 
To ensure the experiment is controlled, we change one hyperparameter at one time, while fixing other hyperparameters.
\begin{itemize}
    \item \textbf{Impact of the parameter $m$.}  The parameter $m$ indicates the number of items that are truncated from the end of the sequence.
    To exploit the impact of the parameter $m$, we vary it in the range of $\{1, 2, 3, 4, 5, 6\}$ and the performance of MSDCCL is shown in Figure \ref{fig:hyperparameter_sensitivity_m}.  It can be observed that on all datasets except Yelp, the model's performance improves with increasing value of $m$ and reaches a peak when $m$ equals to 2 or 3.  This phenomenon occurs because the truncated subsequence contains more enriched short-term interest information of the user compared to the last item in the sequence.  If we continue to raise the value of  $m$, the performance will start to drop. The reason is that when $m$ becomes too large, more outdated items will be considered to represent user preference, which will inevitably leads to inferior performance. On the Yelp dataset, we  observe fluctuations in the model's performance as $m$ increases. This variability may be due to the presence of noise (i.e., misclicks, malicious false interactions) within the truncated subsequence or sudden shifts in the user's short-term interests, resulting in a suboptimal representation of the user's short-term preferences.
   
    \item \textbf{Impact of the parameter $\lambda$.}  We investigate the effect of the hyperparameter $\lambda$  in Eq. (\ref{eq:loss}), and vary it from 0.2 to 2 with a step size of 0.2. The results are reported in Figure \ref{fig:hyperparameter_sensitivity_lambda}. 
    It can be observed that for all datasets, there is a general decrease in model performance with the increasing value of $\lambda$. This observation highlights that it will degrade the performance of sequential recommendations when the Dual Loss with Hard Signal Denoising module takes too much weights in the learning process.
    It is also worth noting that if we set $\lambda$ to 0, which is equivalent to discard the Dual Loss with Hard Signal Denoising module from MSDCCL, the performance will drop significantly. 

\end{itemize}

\section{Conclusion} 
In this paper, we propose a novel denoising framework MSDCCL for sequential recommendation. To be specific, we first employ 
both the soft and hard denoising strategy to alleviate the influence of noisy items in  sequences.  Then, we capture both user long-term and short-term interest via developing a target-aware user interest extractor. Next, we extend existing curriculum learning by simulating the learning pattern of human beings by utilizing the S-shape increment rather than the conventional linear increment. Extensive experiments on five widely used datasets are leveraged for evaluating the performance of the proposed approach. The results show that MSDCCL can significantly boost the performance of existing mainstream sequential recommendation methods. In addition, MSDCCL is also superior to all state-of-the-art denoising models. 
For future work, we will propose to employ data augmentation for addressing the denoising issue. To the end, we introduce an auxiliary task which can provide explicit noisy signals for learning robust denoising models. In addition, we also attempt to investigate other patterns, instead of S-shape increment, for increasing “difficult” instances in curriculum learning.
 
\printcredits

\section*{Acknowledgement}
This work was supported by the National Natural Science Foundation of China [grant number 62141201];  the Natural Science Foundation of Chongqing, China [grant number CSTB2022NSCQ-MSX1672]; the Chongqing Talent Plan Project, China [grant number CSTC2024YCJH-BGZXM0022]; the Major Project of Science and Technology Research Program of Chongqing Education Commission of China [grant number KJZD-M202201102].

\bibliographystyle{cas-model2-names}
 
\bibliography{ref}

\end{document}